\documentclass[10pt]{amsart}

\usepackage[margin=1in]{geometry}
\usepackage{graphicx}
\usepackage{amsmath}
\usepackage{amsxtra}
\usepackage{amstext}
\usepackage{amssymb}
\usepackage{amscd}
\usepackage{amsthm}
\usepackage{url}

\usepackage{algorithmic}
\usepackage{algorithm}

\newtheorem{definition}{Definition}

\newcommand{\VR}{\operatorname{VR}}
\newcommand{\LW}{\operatorname{LW}}
\newcommand{\Ho}{\operatorname{H}}

\title{Applications of Zigzag Persistence to Topological Data Analysis}

\author{Andrew Tausz}
\thanks{The first author was supported by AFOSRG grant FA9550-09-0-1-0531.}
\address{Stanford University, Stanford, CA, 94305}
\email{atausz@stanford.edu}

\author{Gunnar Carlsson}
\thanks{The second author was supported by AFOSRG grant FA9550-09-0-1-0531, ONR grant N00014-08-1-0931 and NSF grant DMS 0905823.}
\address{Stanford University, Stanford, CA, 94305}
\email{gunnar@math.stanford.edu}

\date{\today}

\begin{document}

\maketitle

\begin{abstract}
The theory of zigzag persistence is a substantial extension of persistent homology, and its development has enabled the investigation of several unexplored avenues in the area of topological data analysis. In this paper, we discuss three applications of zigzag persistence: topological bootstrapping, parameter thresholding, and the comparison of witness complexes.
\end{abstract}

\section{Introduction}

The newly emerging area of topological data analysis attempts to use techniques from algebraic topology to study qualitative properties of datasets. Its applicability has been demonstrated in areas as diverse as object recognition, sensor networks, and bioinformatics \cite{Carlsson_09}. The need for a topological approach to data analysis tasks is justified by high-dimensional and highly nonlinear data sources which are not amenable to traditional statistical techniques. One of the key tools in this field is persistent homology, which produces a concise yet meaningful descriptor of an object or point cloud across all spatial scales.

The standard pipeline for performing a persistent analysis of a point cloud consists of two stages:
\begin{enumerate}
\item A filtered simplicial complex is built which consists of nested topological spaces $\{S_r\}$. This filtered complex can be regarded as encoding the mutual connectivity information between all points in the data set. In practice, since we are only interested in finite simplicial or cell complexes, we have a finite sequence
$$S_0 \rightarrow S_1 \rightarrow \ldots \rightarrow S_n$$
where the arrows are inclusions. Standard methods for creating such complexes from geometric data include the Vietoris-Rips, \v{C}ech, $\alpha$-shape and witness constructions.
\item The persistent homology of this filtered complex is computed. In other words, we fix a field, $\mathbb{F}$, and apply the functor $\Ho_p(-, \mathbb{F})$ to the above diagram to obtain the following sequence of finite $\mathbb{F}$-vector spaces
$$\Ho_p(S_0, \mathbb{F}) \rightarrow \Ho_p(S_1, \mathbb{F}) \rightarrow \ldots \rightarrow \Ho_p(S_n, \mathbb{F})$$
The theory of persistent homology, \cite{Carlsson_04}, tells us that such a sequence (which we call a persistence module) can be classified up to isomorphism by a multi-set of intervals which we informally call a barcode. Long intervals indicate the presence of topological features across a wide range of spatial scales. 
\end{enumerate}

The theory of zigzag persistence provides an extension of persistent homology to diagrams of topological spaces of the form:
$$S_0 \leftrightarrow S_1 \leftrightarrow \ldots \leftrightarrow S_n$$
where the arrows can point either left or right. In \cite{ZigZag1} the algebraic foundations are laid out for the theory of zigzag modules which are sequences of vector spaces and linear maps of the form:
$$V_0 \leftrightarrow V_1 \leftrightarrow \ldots \leftrightarrow V_n$$
It is shown that such sequences have the structure of an abelian category and can be classified up to isomorphism by a multi-set of intervals $\{[a_i, b_i] \subset \{0, \ldots, n\}\}$. In \cite{ZigZag2}, an algorithm is presented which allows for the computation of the interval decomposition of the zigzag module:
$$\Ho_p(S_0) \leftrightarrow \Ho_p(S_1) \leftrightarrow \ldots \leftrightarrow \Ho_p(S_n)$$
where $S_i$ are simplicial (or cell) complexes and the arrows are all of the form $S \rightarrow S \cup \{ \sigma \}$ or $S \cup \{ \sigma \} \leftarrow S$. In other words, all of the arrows correspond to the addition or deletion of simplices. The algorithm is essentially an extension of the persistent homology algorithm presented in \cite{Carlsson_04}.

In this paper we explore the use of zigzag homology for studying the topological information contained in point cloud datasets. The following applications were presented (hypothetically) in \cite{Carlsson_09} and will be the object of our study for the remainder of the paper. 

\begin{itemize}
\item {\bf Topological bootstrapping: } Suppose that we are presented with a dataset $X$. We are interested in understanding the homology of the entire dataset from subsamples. It may be the case that $X$ is very large, or that the points in $X$ are accessed in an online manner through some sort of querying process. We would like to understand the homological properties of $X$ through smaller samples $X_0, \ldots, X_n$. If we compute the persistent homology based on the individual samples, $X_i$ and find nontrivial homology classes, we would like to know if these are repeated measurements of the same homology class or different ones. To investigate this we consider the \emph{union sequence} of topological spaces:
$$\ldots \leftarrow X_i \rightarrow X_i \cup X_{i+1} \leftarrow X_{i+1} \rightarrow \ldots$$
If we can somehow determine the continuity of homology classes between the terms in the above sequence, that would tell us whether the samples are measuring the same homological features or not.


\item {\bf Thresholding: } Imagine that in addition to our dataset, $X$, we are provided a parameterized \emph{filter function} which we denote $f(\cdot, \theta): X \rightarrow \mathbb{R}$. Given this function, we may consider filtering our dataset by selecting those points which are in the top $T\%$ ranked by $f(\cdot, \theta)$. An example of this is where $f$ is an empirical density estimator. Although there may be some statistical rules of thumb for selecting the appropriate parameter value, $\theta$, we are interested in understanding how the dataset changes as $\theta$ varies. The trouble is that there may be no natural relationship between the top $T\%$ values for different parameters $\theta$ and $\theta'$. To remedy this, one may consider the sequence
$$\ldots \leftarrow X_f[\theta_i, T] \rightarrow X_f[\theta_i, T] \cup X_f[\theta_{i+1}, T] \leftarrow X_f[\theta_{i+1}, T] \rightarrow \ldots$$
or possibly
$$\ldots \rightarrow X_f[\theta_i, T] \leftarrow X_f[\theta_i, T] \cap X_f[\theta_{i+1}, T] \rightarrow X_f[\theta_{i+1}, T] \leftarrow \ldots$$
where $X_f[\theta, T]$ denotes the top $T\%$ values ranked by $f(\cdot, \theta)$. Computing the homology of the above sequences reveals important information about how the the parameter affects the homological properties of the dataset.

\item {\bf Witness complex comparison: } There are numerous methods for modelling a discrete set of points, $X$, in a metric space by a simplicial complex. Examples include the Vietoris-Rips complex, \v{C}ech complex, and alpha-shape complex. All of these constructions produce a complex with vertex set equal to the original dataset in question. However, there is also a type of complex, namely the witness complex, that allows one to estimate the topological properties of a point cloud by only looking at a subset of the actual points \cite{Witness}. A subset $L \subset X$ is designated as the landmark set, and its points are the vertices in the witness construction which we denote $W(X; L)$. Points in the complement $X \setminus L$ influence the construction of the witness complex, but do not appear in it. This has two main benefits over conventional simplicial complex constructions:
\begin{enumerate}
\item The witness complex produces approximations that are more resilient to random noise.
\item It reduces the computational burden of computing the persistent homology of a large data set.
\end{enumerate}
However, the question arises of how the persistent homology of the approximating witness complex relates to the persistent homology of the entire dataset. We attempt to address this problem in section \ref{witness} of this paper. While it is not possible to answer this question completely without computing the homology for the whole point cloud, we discuss a method for comparing different subsamples. As in the previous two applications, we construct a zigzag diagram of topological spaces
$$\ldots \rightarrow W(X; L_i) \leftarrow W(X; L_i, L_{i+1}) \rightarrow  W(X; L_{i+1}) \leftarrow W(X; L_{i+1}, L_{i+2}) \rightarrow \ldots$$
Homological features that are stable across different landmark selections will manifest themselves as long barcodes in the interval decomposition of the homology of the above sequence.
\end{itemize}


\section{Contributions}

The three examples found in the previous section are stated in \cite{ZigZag1} as potential applications of the theory of zigzag persistence. In this paper we discuss in detail the implementation of these ideas and their performance on actual datasets. To the best of our knowledge, this is the first such discussion of these applications other than in a hypothetical context, and the first implementation. The code for computing zigzag homology is included in the Javaplex software package \cite{javaPlex}. The package is freely available for download at \url{http://code.google.com/p/javaplex}.

\section{Topological Bootstrapping}
\label{resampling}

Suppose that we are presented with a dataset $X$. We are interested in understanding the homology of the entire dataset from samples, and how the samples relate to each other. The approach will be reminiscent of the bootstrap method in statistics (see \cite{bootstrap}), where one obtains additional information about a dataset by performing resampling. The picture to keep in mind is the following:

\vspace{0.5cm}
\hspace{2.6cm}\includegraphics[width=10cm]{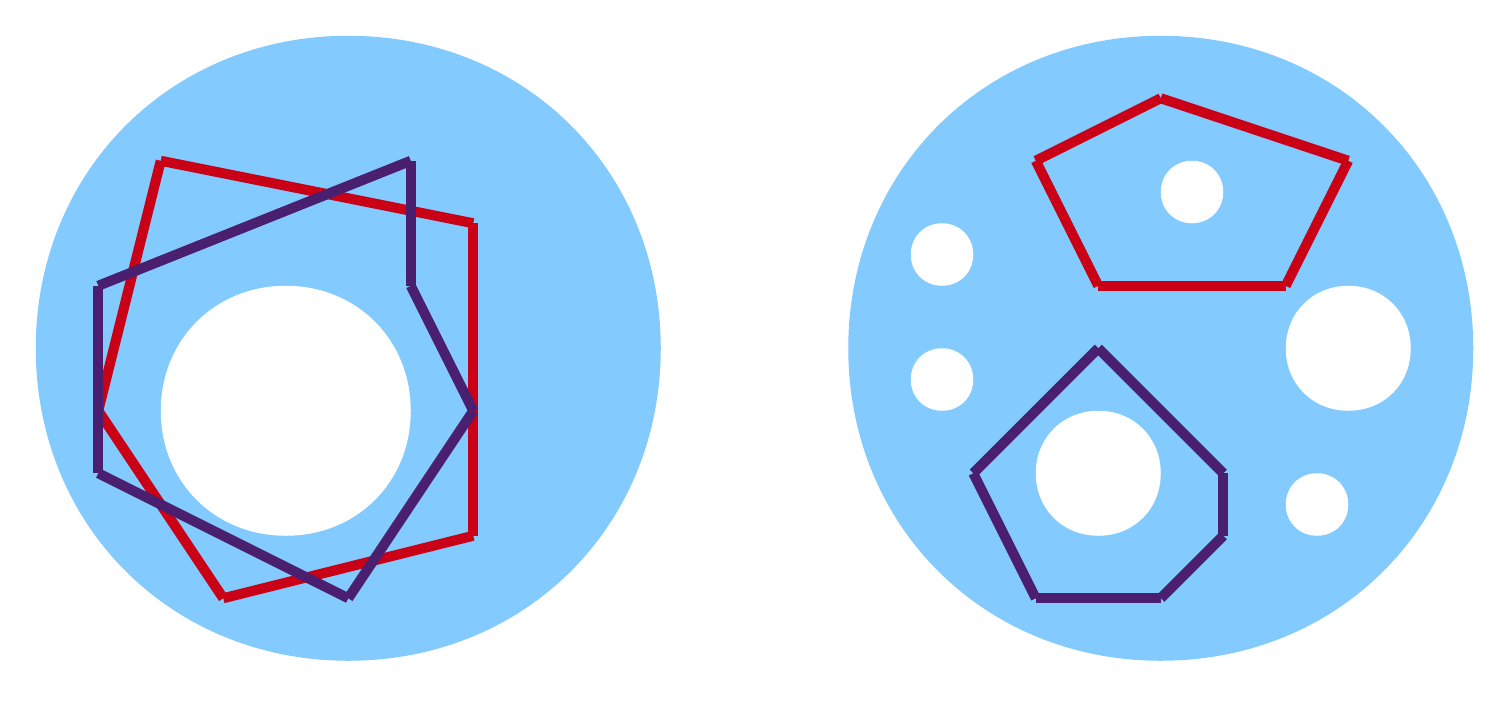}
\vspace{0.5cm}

We are interested to know whether homology classes of the samples are measuring the same homological feature (as on the left) or different features (as on the right). To evaluate the compatibility of two samples $X_i$ and $X_j$ from $X$ we consider their union $X_i \cup X_j$. The Vietoris-Rips construction produces a filtered simplicial complex from a finite metric space as follows: The vertices of $\VR(Y, \epsilon)$ consist of the points of the metric space $Y$. An edge $[u, v]$ is in $\VR(Y, \epsilon)$ if and only if $d(u, v) \leq \epsilon$. For $n > 1$, an $n$-simplex is in the complex if and only if all of its faces are. This complex has the property that:
$$\VR(X_i, \epsilon) \subset \VR(X_i \cup X_j, \epsilon) \supset \VR(X_j, \epsilon)$$
Thus we may consider the zigzag diagram of vector spaces by applying the functor $\Ho_p(-)$ to the above to obtain:
$$\Ho_p(\VR(X_i, \epsilon)) \rightarrow \Ho_p(\VR(X_i \cup X_j, \epsilon)) \leftarrow \Ho_p(\VR(X_j, \epsilon))$$

The interval decomposition of this zigzag diagram tells us important information about the compatibility of the homological features of the complex. If it happens that two classes $[\alpha] \in \Ho_p(\VR(X_i, \epsilon))$ and $[\beta] \in \Ho_p(\VR(X_j, \epsilon))$ map to the same element in $\Ho_p(\VR(X_i \cup X_j, \epsilon))$, this suggests that $[\alpha]$ and $[\beta]$ are measurements of the same $p$-dimensional homological feature.

This idea can be extended to multiple samples $\{X_0, \ldots X_n\}$ from $X$. This provides us with the diagram:
$$ \ldots \rightarrow \VR(X_{i-1} \cup X_{i}, \epsilon) \leftarrow \VR(X_i, \epsilon) \rightarrow \VR(X_i \cup X_{i+1}, \epsilon) \leftarrow \VR(X_{i+1}, \epsilon) \rightarrow \VR(X_{i+1} \cup X_{i+2}, \epsilon) \leftarrow \ldots$$
where the arrows are inclusions.

We also note that this can be repeated with other ``inclusion preserving'' simplicial complex constructions. For example, among the parameterized lazy-witness complexes (see \cite{Witness}), when $\nu = 0$ we have that:
$$\LW_0(X_i, L_i) \subset \LW_0(X_i \cup X_j, L_i \cup L_j)$$
However, as one can see in the paper \cite{Witness} the behavior of the Vietoris-Rips complex and the lazy-witness complex for $\nu = 0$ are very similar, therefore there is not much benefit of one over the other. For $\nu \neq 0$ the construction is not inclusion preserving.

\subsection{Algorithms}

Suppose that we have a point cloud equipped with a metric (in other words, a finite metric space) denoted by $(X, d)$ and subsets $X_0, \ldots, X_n$ with $X_i \subset X$. We also suppose that we have an inclusion preserving filtered complex construction which takes in a finite metric space and outputs a filtered simplicial (or cell) complex, which we denote by $F$. We require $F$ to be inclusion preserving (or functorial in an appropriate sense) so that $X \subset Y \implies F(X) \subset F(Y)$. Note that in practice we most likely use $F = \VR(-, \epsilon)$ to be the Vietoris-Rips complex. In this section we give two algorithms for computing the interval decomposition of the sequence
$$ \ldots \rightarrow \Ho_p(F(X_{i-1} \cup X_{i})) \leftarrow \Ho_p(F(X_i)) \rightarrow \Ho_p(F(X_i \cup X_{i+1})) \leftarrow \Ho_p(F(X_{i+1})) \rightarrow \Ho_p(F(X_{i+1} \cup X_{i+2})) \leftarrow \ldots$$
The first one relies on the incremental addition and removal algorithms presented in section 4 of \cite{ZigZag2}. These algorithms maintain a set of three matrices $Z, B, C$ which essentially store a consistent basis for the vector spaces $\Ho_p(S_i)$. Let us denote by ADD($\sigma$, $k$, $\mathcal{I}$) the algorithm that updates these matrices (the state) with the addition of the simplex $\sigma$ at the index $k$. The set of intervals is denoted by $\mathcal{I}$, and the ADD routine returns a new copy of $\mathcal{I}$ possibly updated with a new interval. Similarly, denote by REMOVE($\sigma$, $k$, $\mathcal{I}$) the algorithm that updates the state with the removal of the simplex $\sigma$ at index $k$. Using these two as subroutines, the interval decomposition of the union zigzag sequence may be computed as follows. Note that we index the terms $F(X_i)$ with $i$ and the terms $F(X_i \cup X_{i+1})$ using $i + \frac{1}{2}$.

\begin{algorithm}
\caption{Interval decomposition of union zigzag (version 1)}
\label{unionalg1}
\begin{algorithmic}


\FOR{$\sigma$ in $F(X_0)$}
\STATE Call $\mathcal{I} \gets$ ADD($\sigma$, 0, $\mathcal{I}$)
\ENDFOR

\FOR{$i = 0, \ldots, n-1$}

\FOR{$\sigma$ in $F(X_i \cup X_{i+1}) \setminus F(X_{i})$}
\STATE Call $\mathcal{I} \gets$ ADD($\sigma$, $i + \frac{1}{2}$, $\mathcal{I}$)
\ENDFOR

\FOR{$\sigma$ in $F(X_{i+1}) \setminus F(X_{i})$}
\STATE Call $\mathcal{I} \gets$ REMOVE($\sigma$, $i + 1$, $\mathcal{I}$)
\ENDFOR

\ENDFOR

\STATE Optionally remove intervals in $\mathcal{I}$ supported on half-integral indices.

\end{algorithmic}
\end{algorithm}

Despite the simplicity of the above algorithm, it is also possible to approach the task slightly differently. If we compute the persistent homology of $F(X_j)$, $F(X_{j+1})$ and $F(X_j \cup X_{j+1})$, we are interested in the induced action on homology on the inclusion maps $\imath: X_j \hookrightarrow X_j \cup X_{j+1}$. However, the only two possibilities are that $\imath_*$ is the identity on a homology class $[a] \in \Ho_p(F(X_j))$ or is the zero map on $[a]$. Thus since we are given the computed persistent homology of $F(X_j \cup X_{j+1})$, we can ``match'' homology classes. Note that the persistent homology algorithm (as well as the algorithm in \cite{ZigZag2}) attempts to perform the computation incrementally by computing the persistent homology of $K \cup \{\sigma\}$ from $K$. However, in this situation the incremental approach is not necessary: We may compute the persistent homology of $F(X_j)$, $F(X_{j+1})$ and $F(X_j \cup X_{j+1})$ and simply examine which classes are preserved or killed-off by the induced inclusion map. We present this algorithm below for the union sequence. The algorithm maintains a collection of intervals, which at the $j$-th stage of the algorithm is denoted by $\mathcal{I}_j$. The completed interval decomposition is given by $\mathcal{I}_n$.

\begin{algorithm}
\caption{Interval decomposition of union zigzag (version 2)}
\label{unionalg2}
\begin{algorithmic}

\STATE Initialize $\mathcal{I}_0 \gets \{I_a = [0, \infty] : \mbox{ $a$ is an active homology class in $\Ho_p(F(X_0))$} \}$

\FOR{$j = 0, \ldots, n-1$}

\STATE $\mathcal{I}_j \gets \mathcal{I}_{j-1}$

\FOR{$[a] \in \Ho_p(F(X_j))$} 
\FOR{$[b] \in \Ho_p(F(X_{j+1}))$} 

\IF {$[a]$ and $[b]$ map to the same homology class in $\Ho_p(F(X_j \cup X_{j+1}))$}
\STATE Maintain the interval $I_a = [s_a, \infty]$ corresponding to the homology class $a$ in $\mathcal{I}_j$
\STATE Mark homology classes $[a]$ and $[b]$ as matched
\ENDIF

\ENDFOR
\ENDFOR

\FOR{$[a] \in \Ho_p(F(X_j)) \mbox{ such that $[a]$ is unmatched}$}
\STATE End the interval corresponding to the class $[a]$ by changing $I_a = [s_a, \infty]$ to $I_a \gets [s_a, j]$
\ENDFOR

\FOR{$[b] \in \Ho_p(F(X_{j+1})) \mbox{ such that $[b]$ is unmatched}$} 
\STATE Start an interval corresponding to $[b]$ by adding $I_b = [j, \infty]$ to $\mathcal{I}_j$
\ENDFOR

\ENDFOR

\STATE Close off all remaining intervals of the form $I_a = [s_a, \infty]$ by setting them to $I \gets [s_a, n]$

\end{algorithmic}
\end{algorithm}

There is one drawback to algorithm \ref{unionalg2} in comparison with algorithm \ref{unionalg1}. Algorithm \ref{unionalg2} must compute the persistent homology of the individual terms (such as $F(X_j)$) by calling the ADD routine. Thus, for the sequence $F(X_j) \rightarrow F(X_j \cup X_{j+1}) \leftarrow F(X_{j+1})$, it must call ADD a total of $|F(X_j)| +  |F(X_j \cup X_{j+1})| + |F(X_{j+1})|$ times. In comparison, algorithm \ref{unionalg1} must call ADD $|F(X_j)| +  |F(X_j \cup X_{j+1}) \setminus F(X_j)|$ times, and REMOVE $|F(X_j \cup X_{j+1}) \setminus F(X_{j+1})|$ times. Thus in terms of the total number of ADD and REMOVE calls, algorithm \ref{unionalg2} is suboptimal in comparison with algorithm \ref{unionalg1}, which performs the exact minimal number of ADD and REMOVE operations. However, the difference is that the second algorithm only requires the ADD routine. This has the advantage in that the data structures needed to support the REMOVE operation are more complex and require more overhead than those needed to support only the ADD operation. In our software implementation, we used algorithm \ref{unionalg2}.

Similarly, we may also use the ADD and REMOVE routines to compute the interval decomposition of the intersection sequence
$$ \ldots \leftarrow \Ho_p(F(X_{i-1} \cap X_{i})) \rightarrow \Ho_p(F(X_i)) \leftarrow \Ho_p(F(X_i \cap X_{i+1})) \rightarrow \Ho_p(F(X_{i+1})) \leftarrow \Ho_p(F(X_{i+1} \cap X_{i+2})) \rightarrow \ldots$$
just as easily. For example, the adaptation of algorithm \ref{unionalg1} is shown below. Additionally, one may also adapt algorithm \ref{unionalg2} to the intersection sequence trivially.

\begin{algorithm}
\caption{Interval decomposition of intersection zigzag}
\label{intersectionalg}
\begin{algorithmic}


\FOR{$\sigma$ in $F(X_0)$}
\STATE Call $\mathcal{I} \gets$ ADD($\sigma$, 0, $\mathcal{I}$)
\ENDFOR

\FOR{$i = 0, \ldots, n-1$}

\FOR{$\sigma$ in $F(X_{i}) \setminus F(X_i \cap X_{i+1})$}
\STATE Call $\mathcal{I} \gets$ REMOVE($\sigma$, $i + \frac{1}{2}$, $\mathcal{I}$)
\ENDFOR

\FOR{$\sigma$ in $F(X_{i+1}) \setminus F(X_i \cap X_{i+1})$}
\STATE Call $\mathcal{I} \gets$ ADD($\sigma$, $i + 1$, $\mathcal{I}$)
\ENDFOR

\ENDFOR

\STATE Optionally remove intervals in $\mathcal{I}$ supported on half-integral indices.

\end{algorithmic}
\end{algorithm}

\subsection{Basic Example}

Let us verify our intuition on an example that is easy to verify. The following example shows 7 random samples from a dataset containing 10,000 points on a figure-8. Below, we show the Vietoris-Rips complexes constructed from each sample $X_i$ for $i = 0, \ldots, 6$. A maximum filtration value of 1.2 was used and each sample contains 40 points.

\vspace{0.5cm}
\hspace{-1.5cm}\includegraphics[height=2cm]{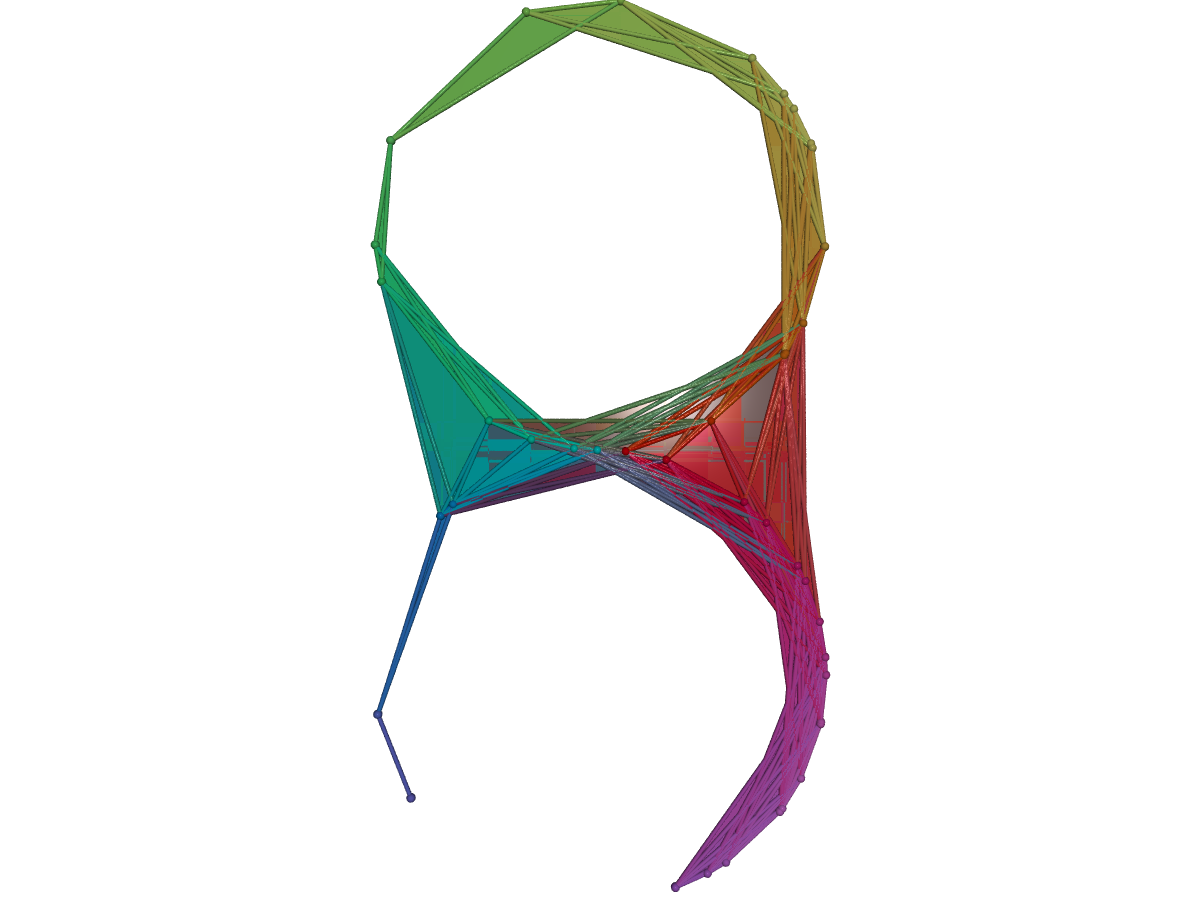}\includegraphics[height=2cm]{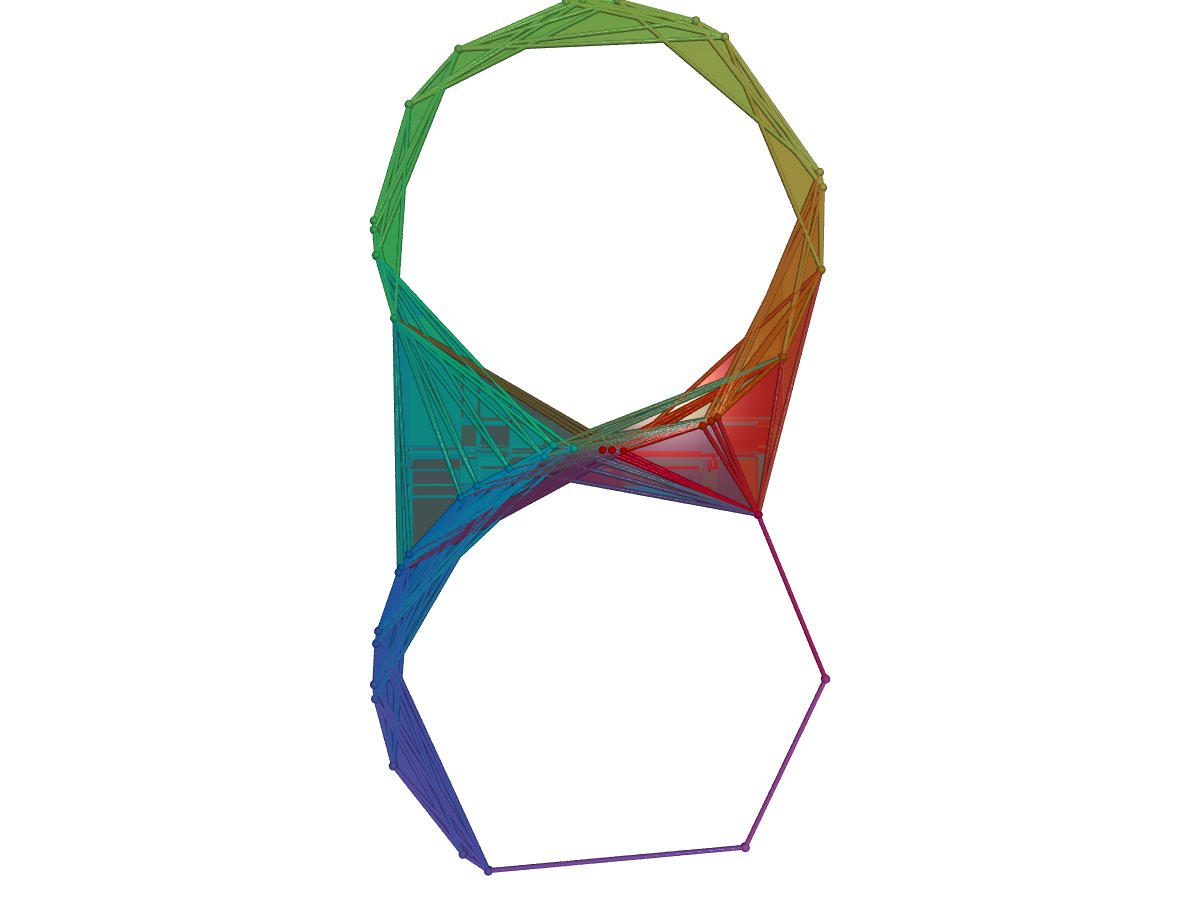}\includegraphics[height=2cm]{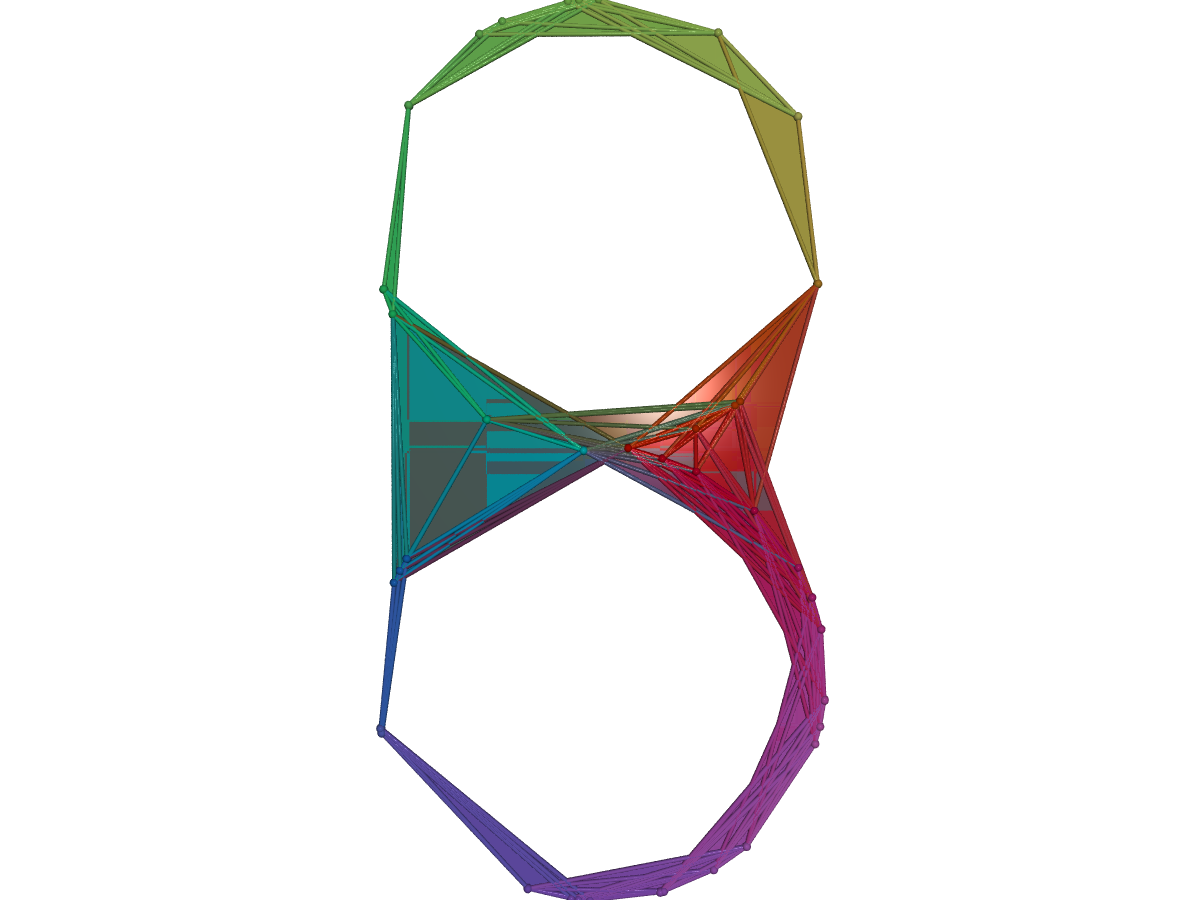}\includegraphics[height=2cm]{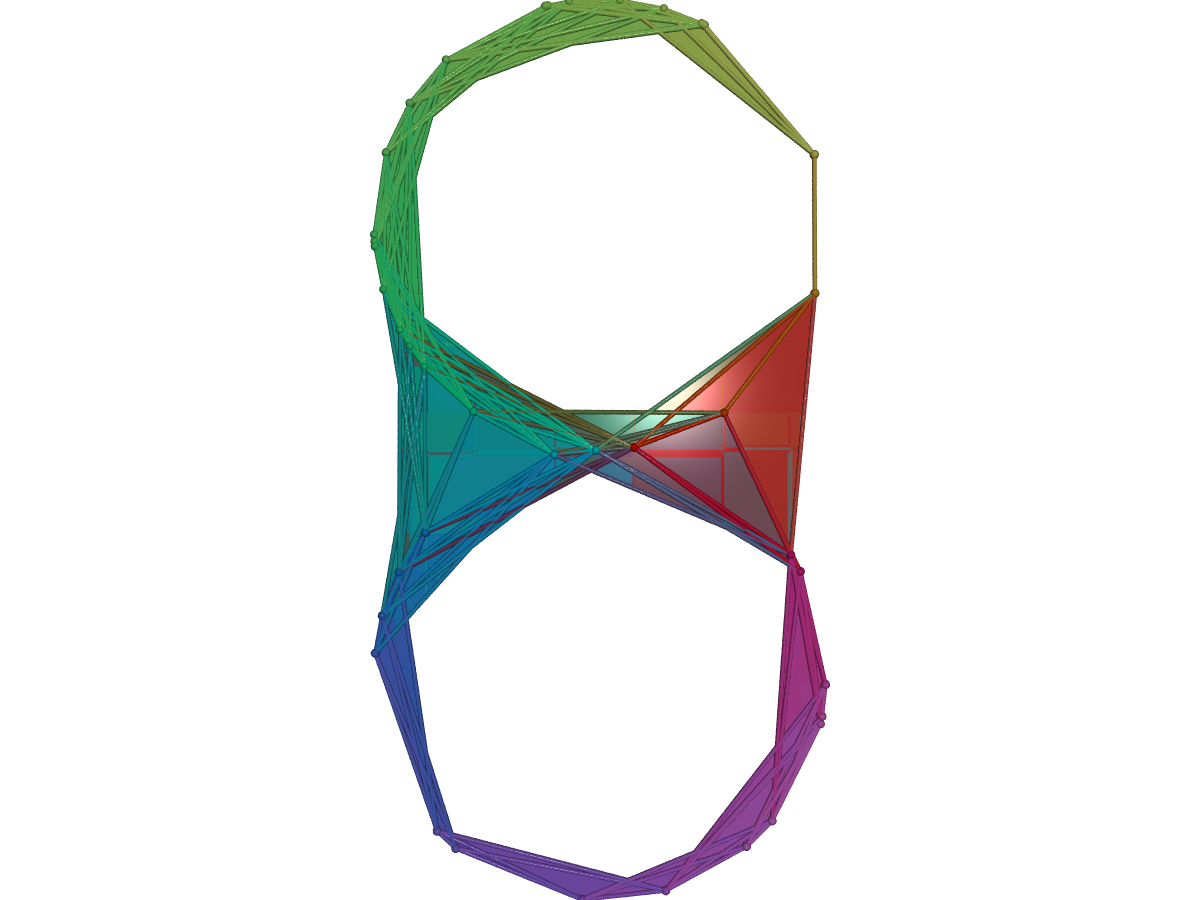}\includegraphics[height=2cm]{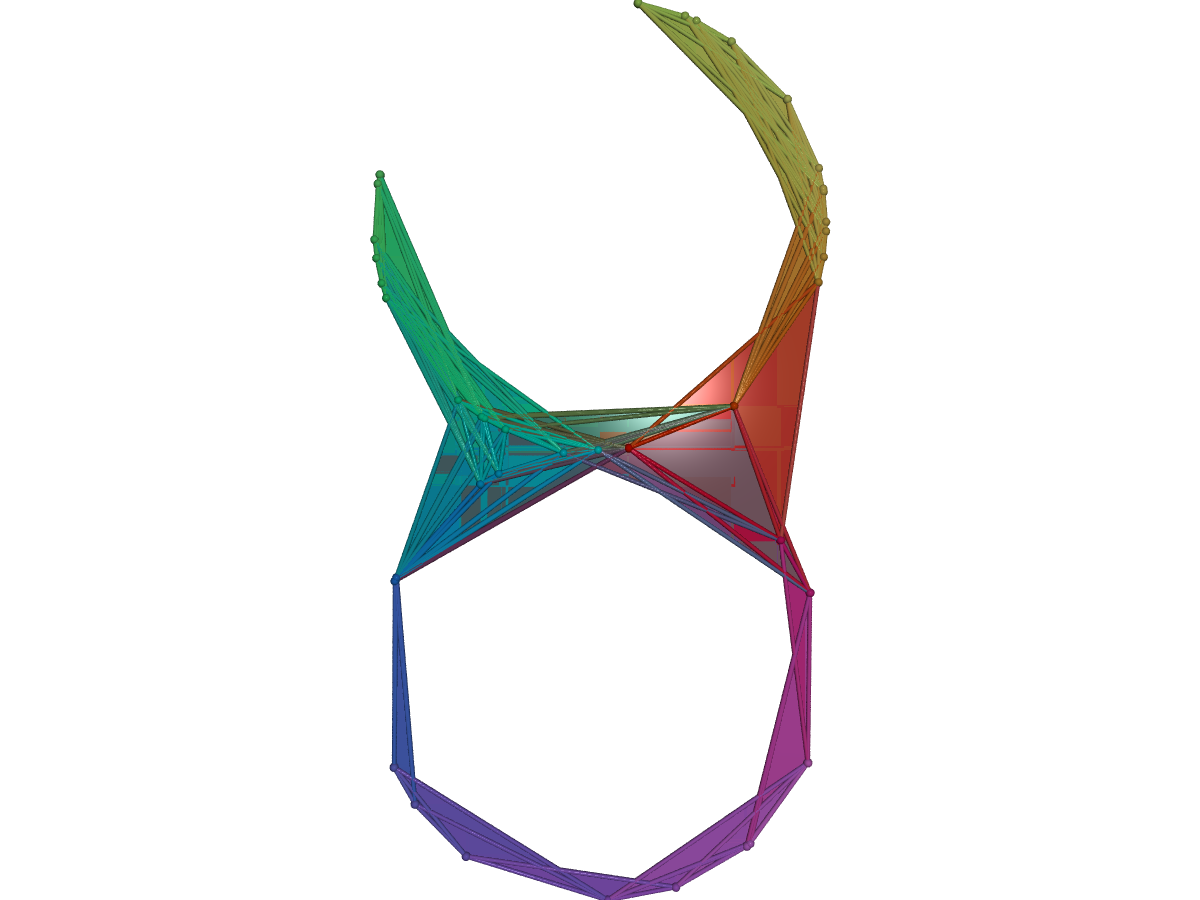}\includegraphics[height=2cm]{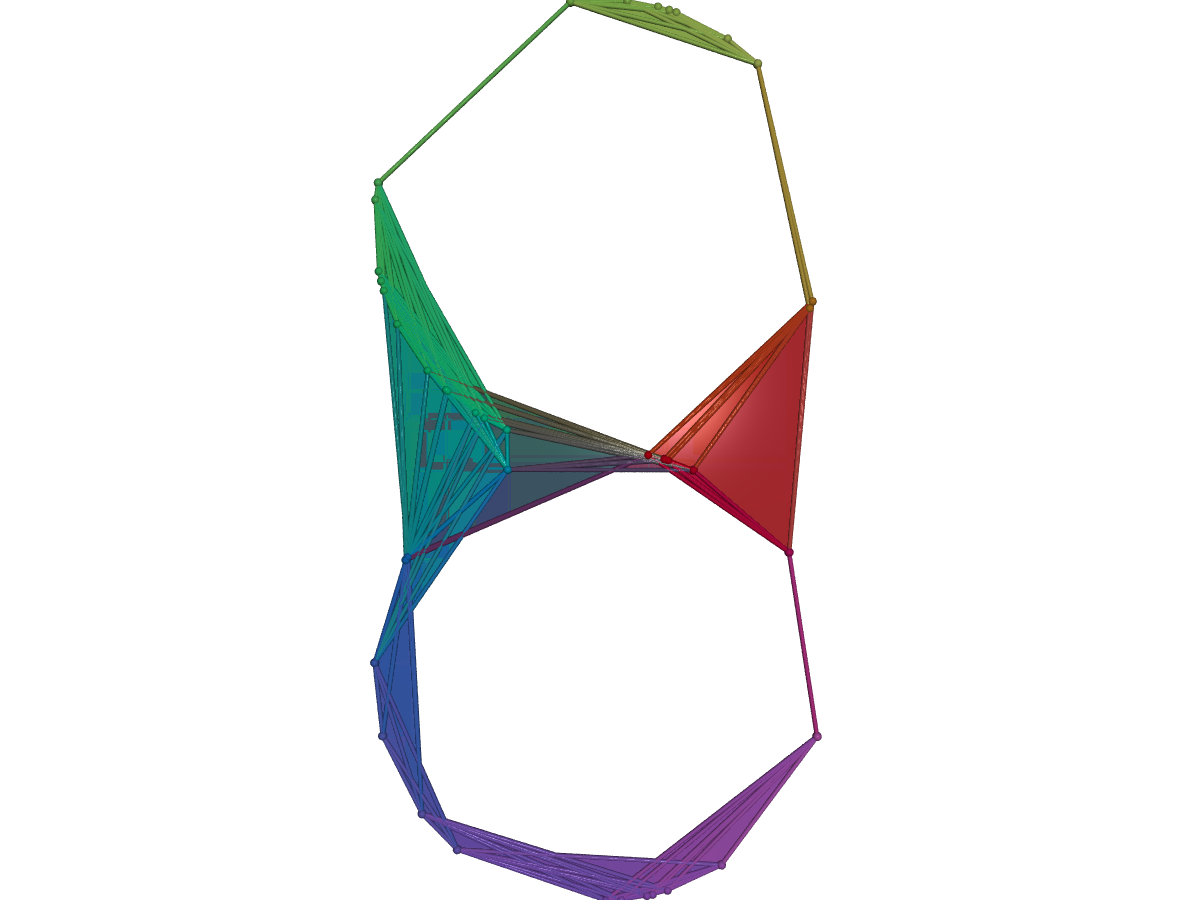}\includegraphics[height=2cm]{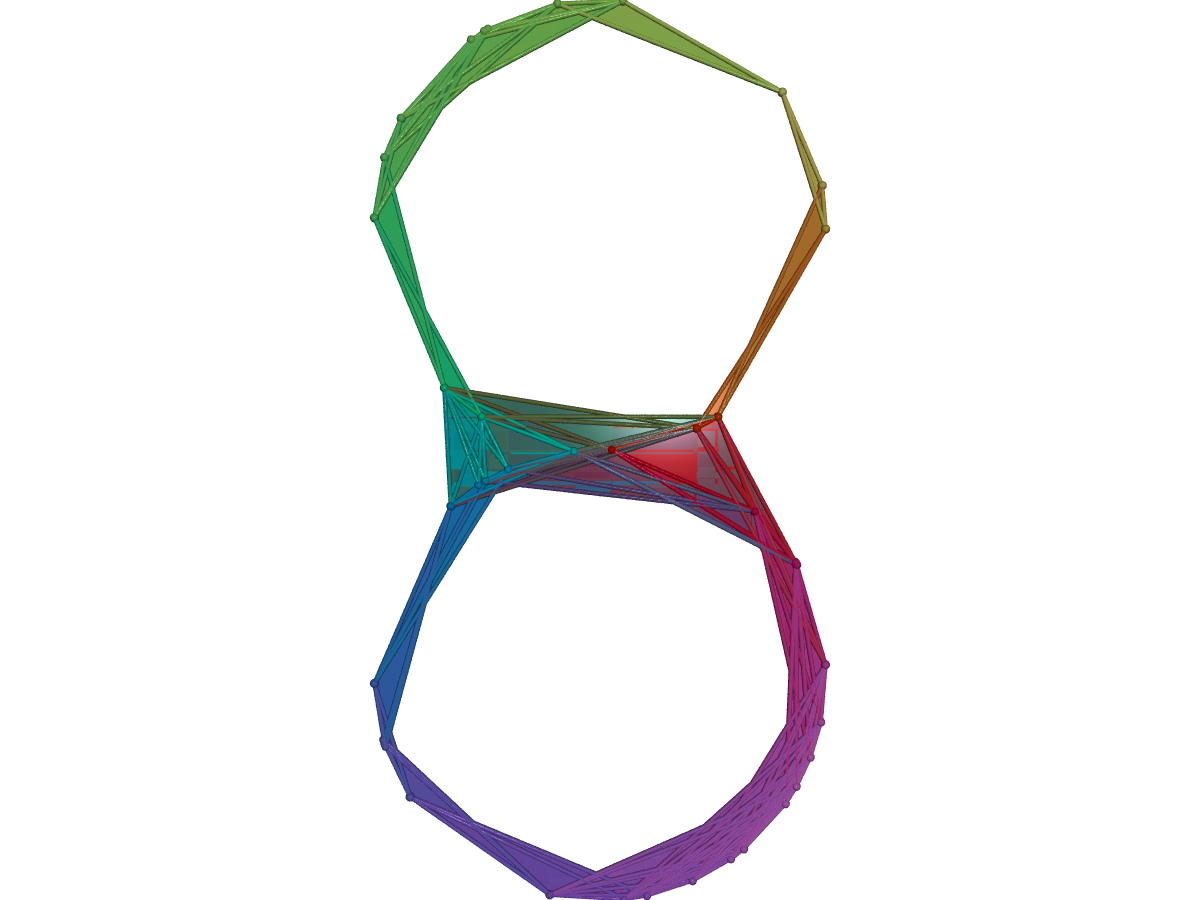}
\vspace{0.5cm}

If we assign indices $X_i \leftrightarrow i$, and $X_i \cup X_{i+1} = X_{i, i+1} \leftrightarrow i + \frac{1}{2}$, and we suppress the homology classes in the intermediate complexes (the unions), then the intervals tell us about the ``transfer'' of homology cycles between the different samples. For the above samples, when one uses the union sequence, the following intervals are obtained:

\vspace{0.5cm}
\hspace{3cm}\includegraphics[width=10cm]{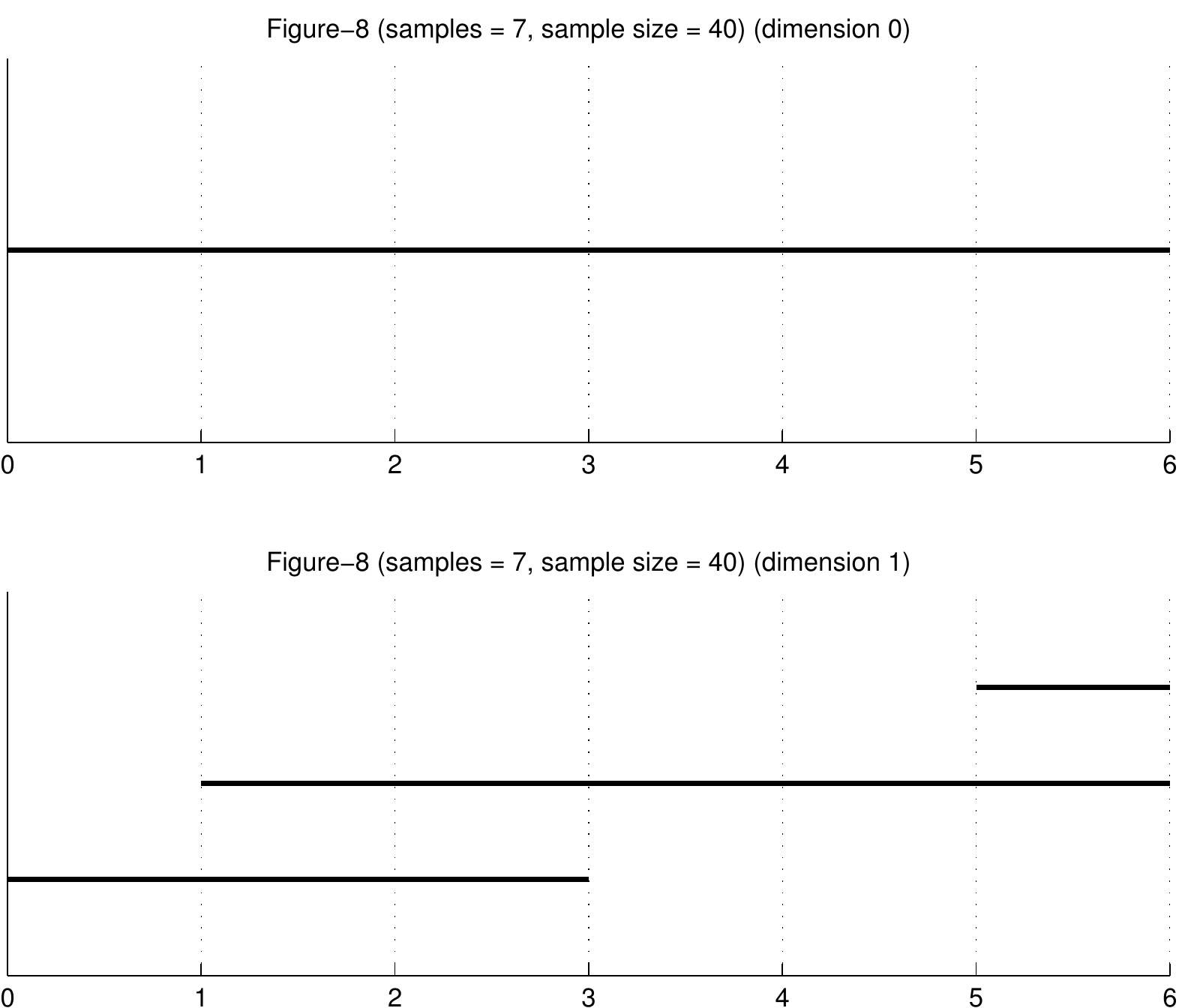}
\vspace{0.5cm}

The cardinality of the lines above a point on the horizontal axis indicates the rank of the corresponding homology group. For example, for the sample at index 0, we can see that the homology groups both have rank 1 in dimension 0 and 1. At index 1, they have ranks 1 and 2 in dimensions 0 and 1, respectively. Furthermore, we also see that the 1-dimensional cycle computed at index 0 is compatible with one of the two measured at index 1, due to the continuity of the interval. Similarly, we can see that all of the 0-dimensional homology classes are the same (in other words, they represent the same connected component). In indices 1-3, we can see that the two homology classes measured in dimension 1 are also pairwise compatible. 

We do not consider the intersection sequence with intermediate terms $X_i \cap X_{i+1}$ here since they intersect extremely sparsely for random samples. 

Similarly, the following figure shows the interval decomposition of the union sequence where the underlying space is a random sample of 10,000 points on the unit 2-sphere. 10 samples of 100 points were taken and a maximum filtration value of 1.0 was used for constructing the Vietoris-Rips complexes. Note that at indices 3 and 6 the random subsample did not find a 2-dimensional homology class. However, for neighboring indices which did find a 2-dimensional class the barcode plot shows that these classes are compatible.

\vspace{0.5cm}
\hspace{3cm}\includegraphics[width=10cm]{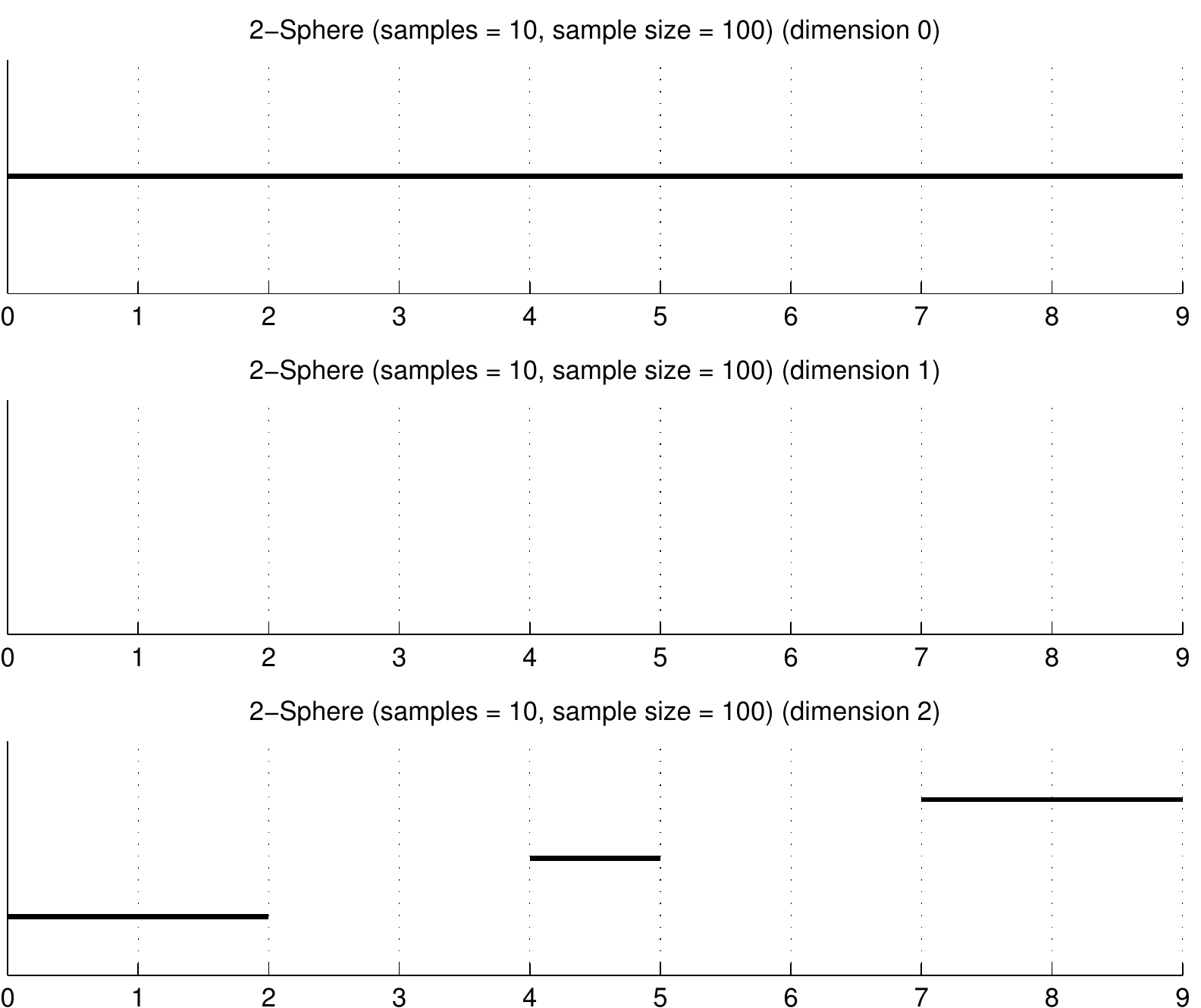}
\vspace{0.5cm}

\subsection{Incremental Samples}

Using the Vietoris-Rips based sampling framework, we may also investigate the role of the sample size. To do this, we generate samples $\{X_0, \ldots, X_n\}$ with $|X_i| = N_i$. In the example below we perform this from a dataset which consists of 10,000 random points on a figure-8. Samples were generated of sizes $2, 3, 4, \ldots, 150, 151, 152$. A maximum filtration value of 1.0 was used for the construction of the Vietoris-Rips complexes. The following figure shows the barcodes for the homology of this sequence.

\vspace{0.5cm}
\hspace{3cm}\includegraphics[width=10cm]{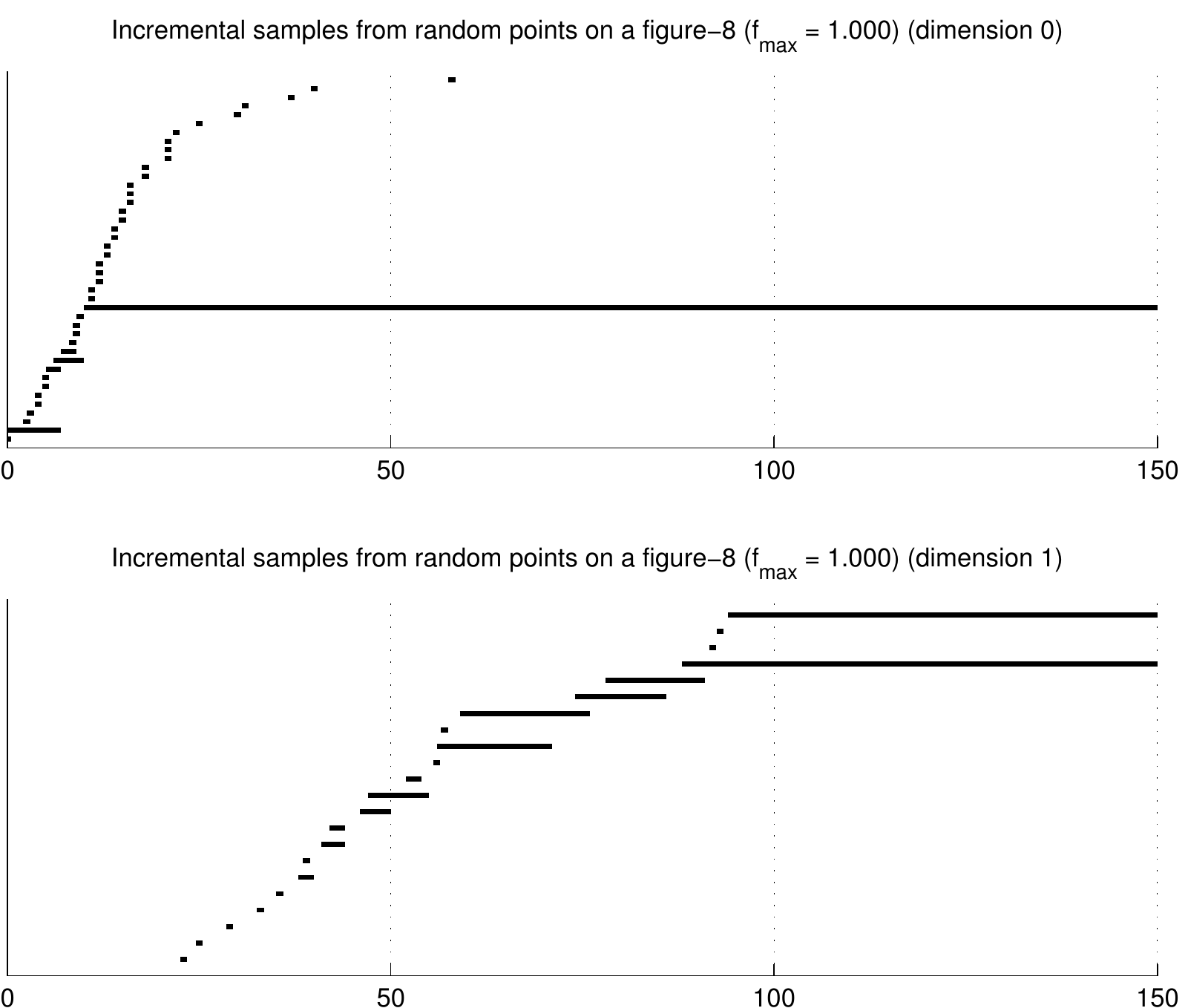}
\vspace{0.5cm}

These barcodes indicate that once the size of the sample is roughly above 100, the correct Betti numbers $\{1, 2\}$ are computed from the samples, and furthermore these homology classes are compatible. It is important to note that the samples are not nested, but are independent uniformly random selections.

\section{Parameterized Filtration}
\label{parameterized}

Suppose that we have a dataset $X$ and a parameterized filtration function $f(\cdot, \theta):X \rightarrow \mathbb{R}$. An example could be a density estimator which is parameterized by some sort of width or variance parameter. One is interested in studying homologically how the filtration function behaves on the dataset for different parameter values. Let us define the set:
$$X_f[\theta, T] = \{x \in X | \mbox{ $x$ is among the top $T\%$ points ranked by the filtration function $f(\cdot, \theta)$}\}$$
We are interested in how the samples relate to each other as we change the parameter $\theta$, thus we consider the sequence of samples for a sequence of parameters $\{\theta_i\}$:
$$\ldots \leftarrow X_f[\theta_i, T] \rightarrow X_f[\theta_i, T] \cup X_f[\theta_{i+1}, T] \leftarrow X_f[\theta_{i+1}, T] \rightarrow \ldots$$
As done previously, we may apply an inclusion preserving filtered complex construction and compute its zigzag homology. For example, we may compute barcodes for:
\begin{equation}
\label{homologysequence}
\ldots \leftarrow \Ho_p(\VR(X_f[\theta_i, T])) \rightarrow \Ho_p(\VR(X_f[\theta_i, T] \cup X_f[\theta_{i+1}, T])) \leftarrow \Ho_p(\VR(X_f[\theta_{i+1}, T])) \rightarrow \ldots
\end{equation}
The existence of intervals of positive length suggest the preservation of homological features across several values of the parameter $\theta$. Similarly, we may also consider the intersection sequence analogous to the above:
\begin{equation}
\label{homologysequence2}
\ldots \rightarrow \Ho_p(\VR(X_f[\theta_i, T])) \leftarrow \Ho_p(\VR(X_f[\theta_i, T] \cap X_f[\theta_{i+1}, T])) \rightarrow \Ho_p(\VR(X_f[\theta_{i+1}, T])) \leftarrow \ldots
\end{equation}

We note that these sequences are no different than those in the previous section. For example, one may write $X_i = X_f[\theta_i, T]$ and then the union and intersection sequences are the same as those in section \ref{resampling}, thus the same algorithms apply. 

\subsection{Image Patch Data Example}
\label{impatch1}

In \cite{Images} and \cite{Carlsson_09}, the authors discuss a topological analysis of a dataset consisting of high-contrast patches from a database of natural images. Using a density filtration, they conclude that a set of such patches has the topology of a Klein bottle. The dataset under consideration, called $\mathcal{M}$, consists of around $4.5 \times 10^6$ patches of $3 \times 3$ pixels appropriately normalized. From this a random sample of $5 \times 10^4$ points is selected which we call $\mathcal{M}_0$. We refer the reader to \cite{Lee} for additional details on the source of the data and the preprocessing steps taken.

In this situation one possible filter function is the $k$-codensity function defined by:
$$\delta(x, k) = \delta_k(x) = d(x, \nu_k(x))$$
where $\nu_k(x)$ denotes the $k$-th nearest neighbor to the point $x$. We may think of $\delta_k(x)$ as being inversely related to the density of $X$ at $x$. The parameter $k$ is a smoothing parameter -- large values of $k$ mean that the codensity is measured in a large region around $x$. As before, we define the sets:
$$X_{\delta}[k, T] = \{x \in X | \mbox{ $x$ is among the $T\%$ densest points in $X$ as ranked by the codensity $\delta_k$}\}$$
Alternatively, we may use a continuously parameterized density estimator:
$$\hat{f}(x, \sigma) = \frac{1}{n} \sum_{i=1}^n K_{\sigma}(x - x_i)$$
where 
$$K_{\sigma}(z) = (\sqrt{2 \pi} \sigma)^{-d} \exp \left( -\frac{||z||^2}{2 \sigma^2} \right)$$
is a circular Gaussian kernel. It turns out that the codensity function $\delta$ and the kernel density function $\hat{f}$ both give the same qualitative conclusions. Our goal is to study the sets$\mathcal{M}_{0, \delta}[k, T]$ or $\mathcal{M}_{0, \hat{f}}[\sigma, T]$ which for notational convenience we denote by $\mathcal{M}_{0}[k, T]$ and $\mathcal{M}_{0}[\sigma, T]$ respectively. 

In \cite{Images} it is shown that when one filters by $\delta$ and sets $k = 300$ one may recover the ``primary circle'' which can be thought of as containing edges of all orientations. The patches on this circle component can be schematically visualized as follows:

\vspace{0.5cm}
\hspace{5cm}\includegraphics[width=6cm]{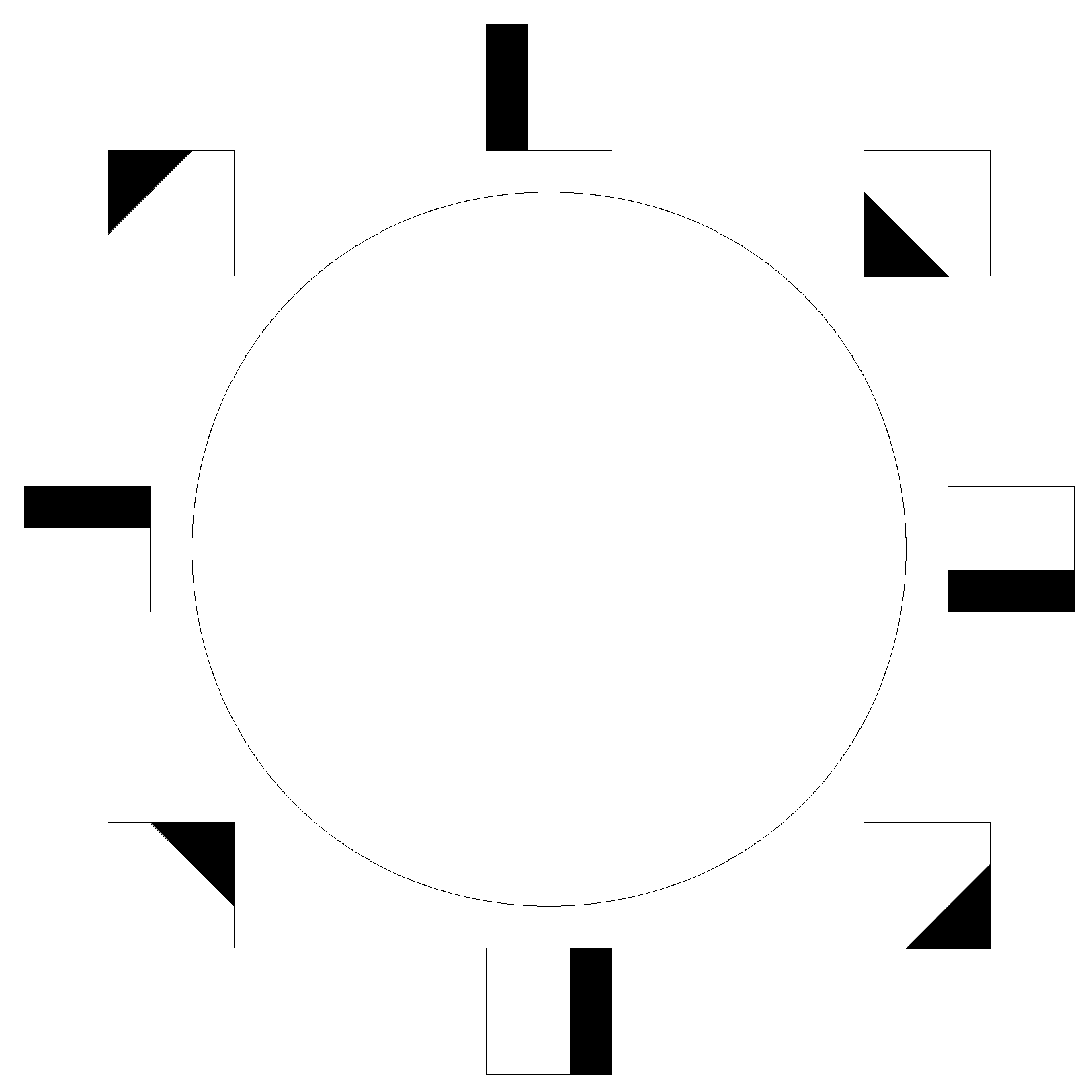}
\vspace{0.5cm}

When using $k = 15$ one obtains the 3-circle model, where in addition to the primary circle there are two other secondary circles. However, in \cite{Images} and \cite{Carlsson_09} these results are obtained by selecting a landmark set and then computing the persistent homology of a lazy-witness complex, which drastically reduces the computational burden. In this case we chose not to do that since we are interested in the compatibility between Vietoris-Rips samples. To illustrate the difficulty, for the image patch data we are interested in the case where $T = 30\%$, so that we have a sample of $1.5 \times 10^4$ points. Constructing a Vietoris-Rips complex on a dataset of this size is well outside the capability of any known software package for computing persistent homology. To remedy this, we do the following. Given $\mathcal{M}_{0}[k, T]$ for a specified value of $k$ and $T$, we choose a subsample of $S$ points via a sequential max-min procedure. The Vietoris-Rips complex is then construct from this subsample of size $S$. In practice, $S$ will be on the order of 100. This procedure is similar to computing a (lazy)-witness complex, except that we ensure that the construction is inclusion preserving.

Taking $k = 300$, $T = 30\%$, and $S = 100$ we observe the primary circle very clearly:

\vspace{0.5cm}
\hspace{3cm}\includegraphics[width=10cm]{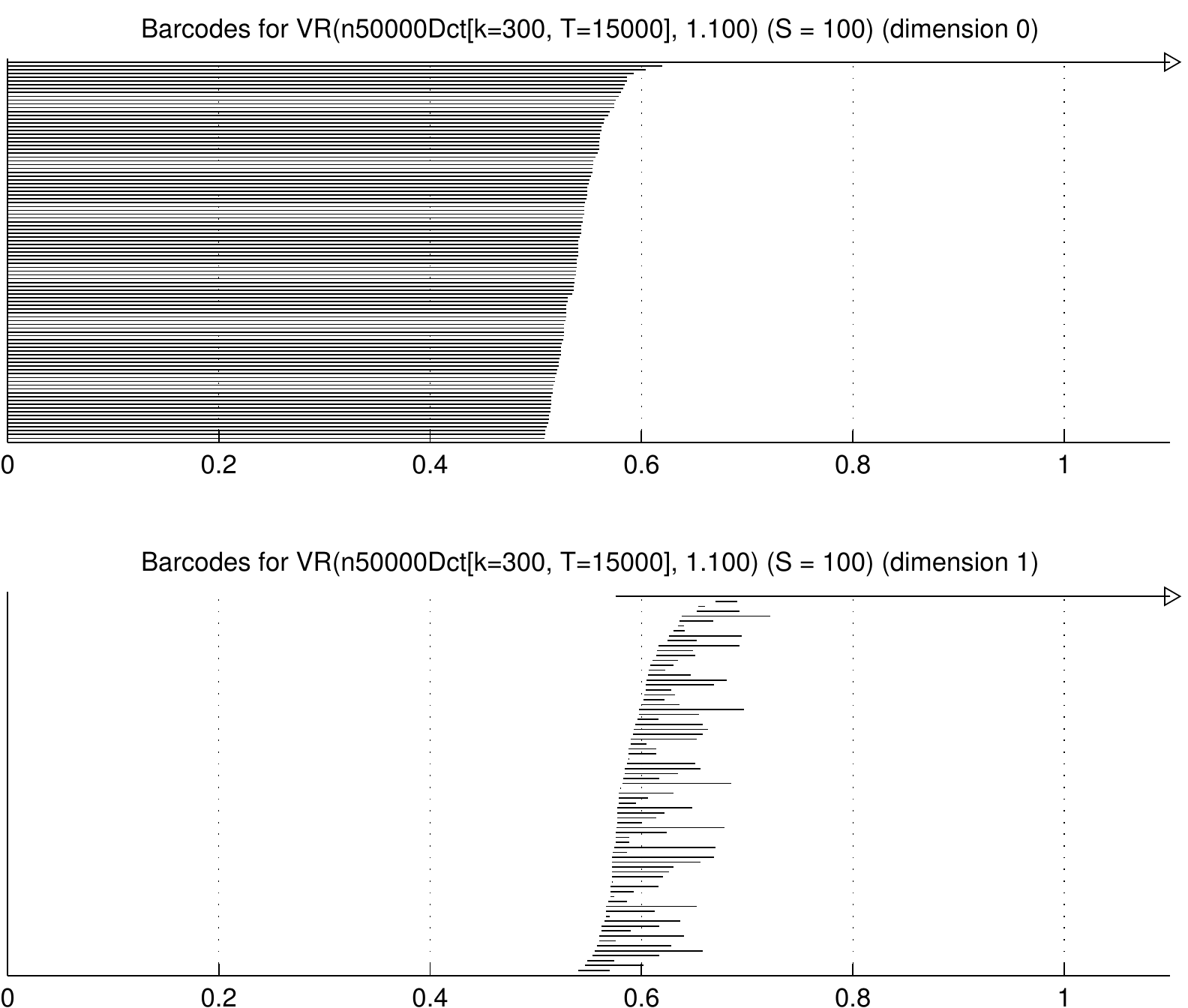}
\vspace{0.5cm}

Taking $k = 15$, $T = 30\%$, and $S = 100$ we can recover the 3-circle model, manifested by the fact that $\beta_1 = 5$.

\vspace{0.5cm}
\hspace{3cm}\includegraphics[width=10cm]{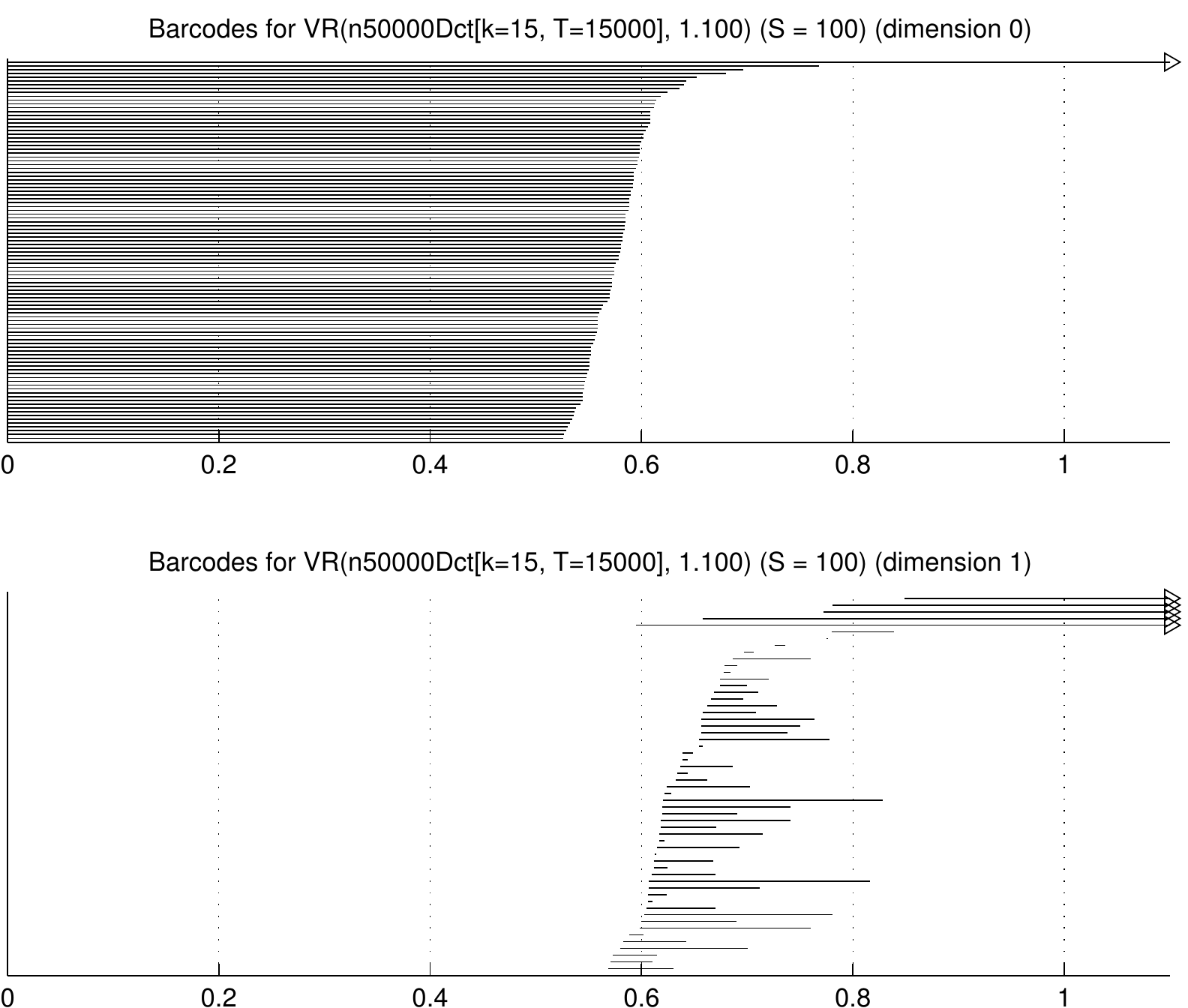}
\vspace{0.5cm}

Similarly, when filtering by $\hat{f}$ one obtains the three circle model for low values of $\sigma$ and the primary circle for higher values. For example, when $\sigma = 0.11$, we observe:

\vspace{0.5cm}
\hspace{3cm}\includegraphics[width=10cm]{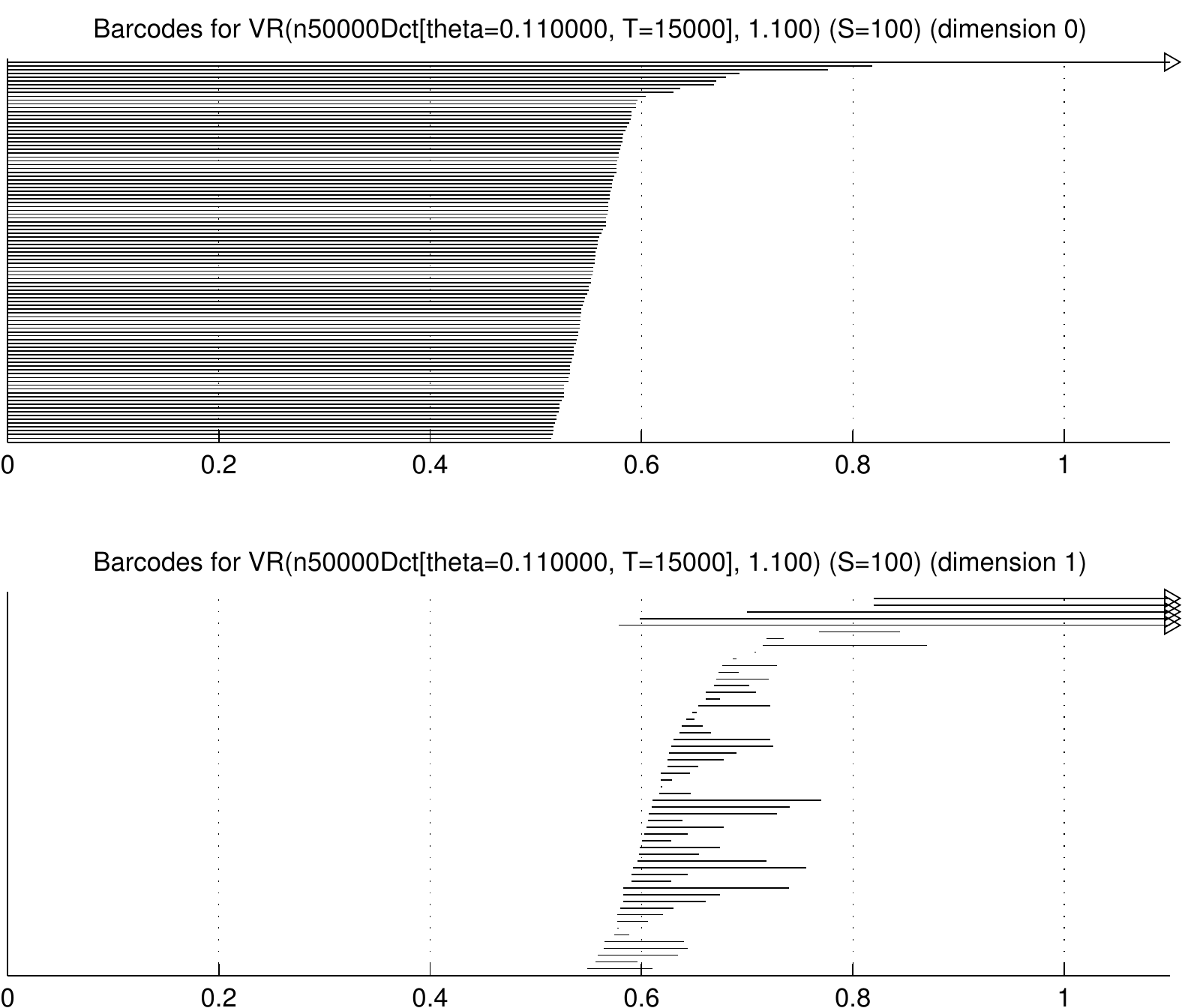}
\vspace{0.5cm}

and when we set $\sigma = 0.2$, the secondary circles disappear:

\vspace{0.5cm}
\hspace{3cm}\includegraphics[width=10cm]{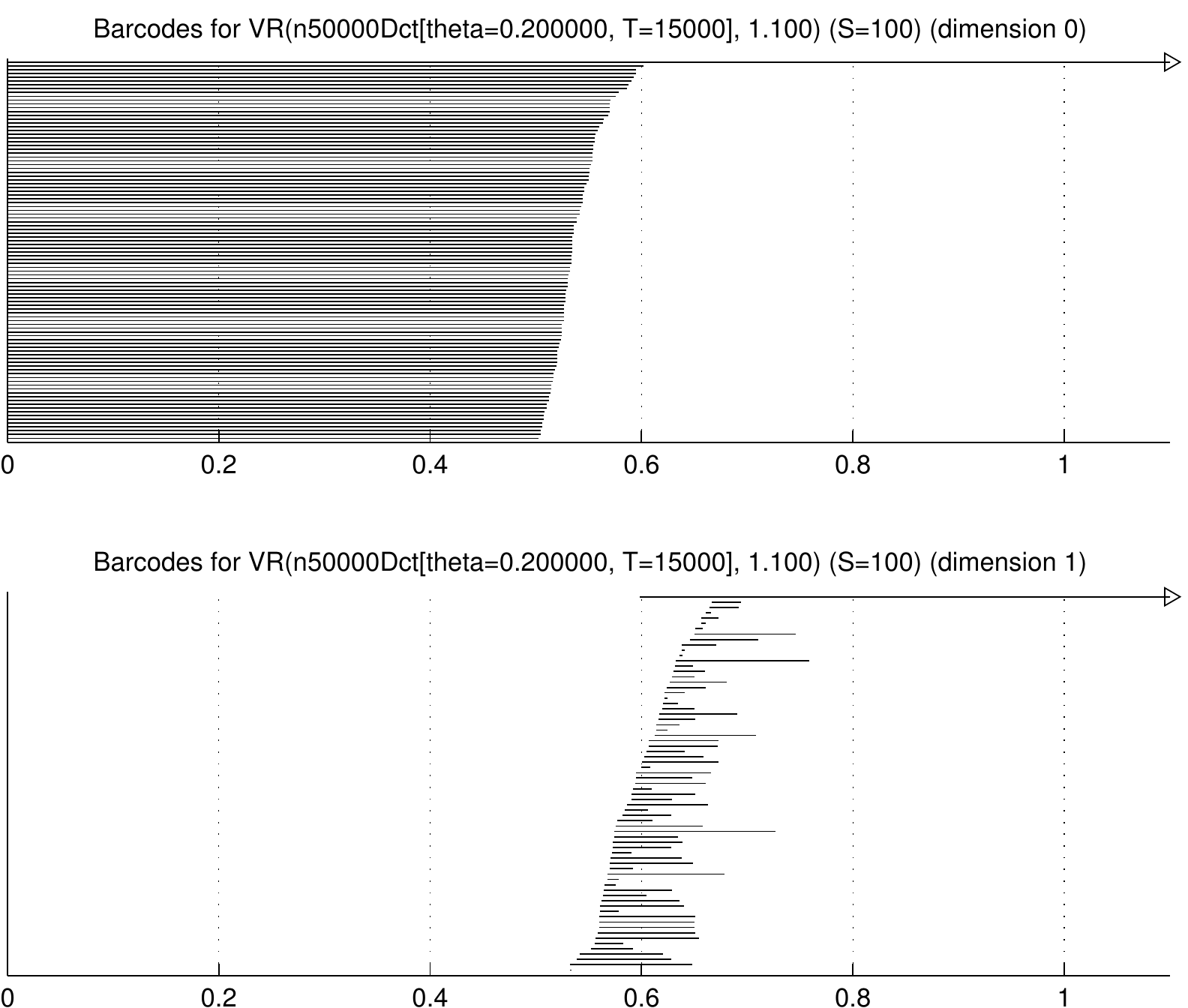}
\vspace{0.5cm}

To see the effect of the density parameter, we compute barcodes for the zigzag sequence (\ref{homologysequence}) using the kernel density estimate $\hat{f}$. In the figure below, this is done for $\sigma = 0.11, \ldots, 0.2$, $T = 30\%$, and $S = 100$. 

\vspace{0.5cm}
\hspace{3cm}\includegraphics[width=10cm]{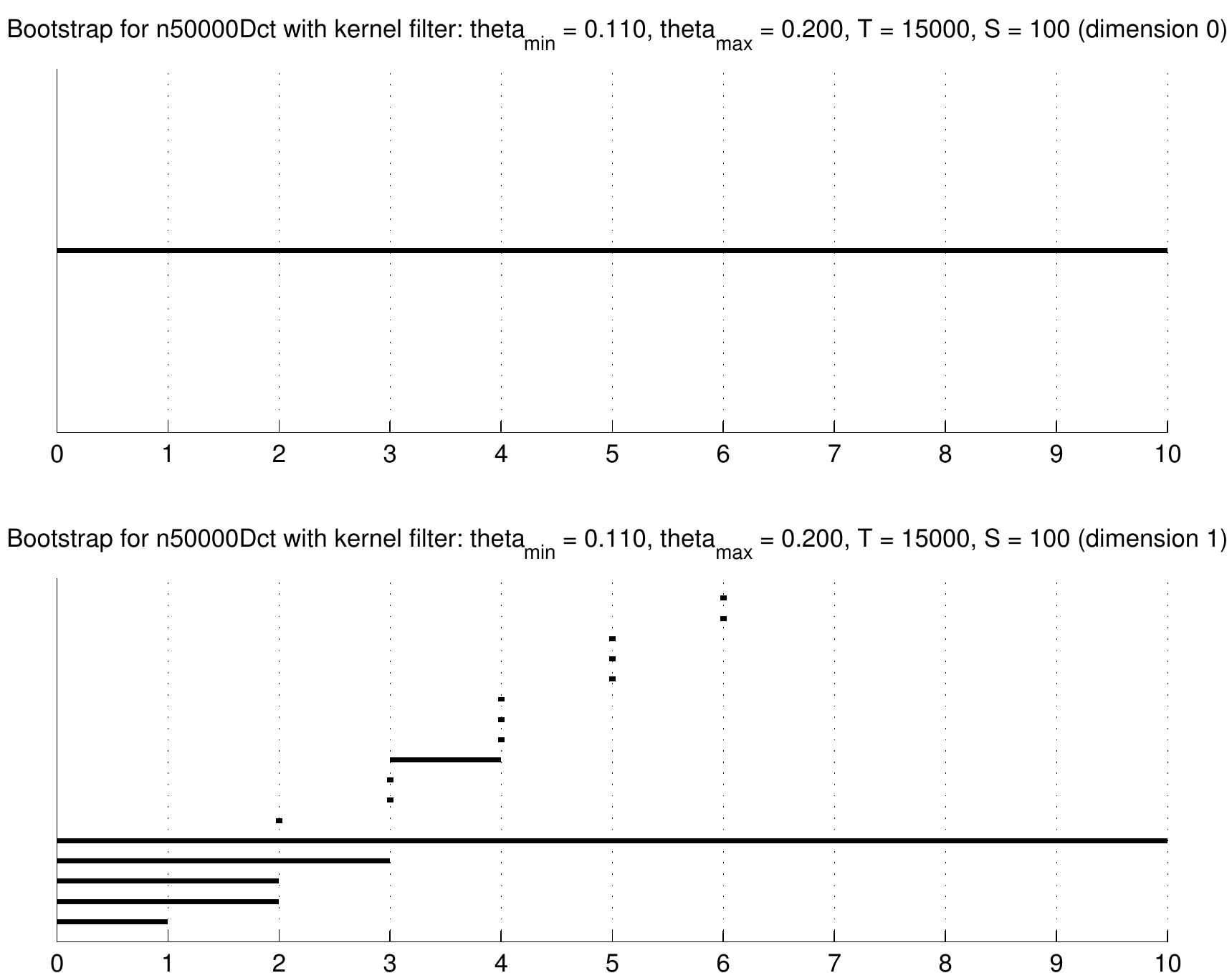}
\vspace{0.5cm}

It should be noted that the ticks on the horizontal axis indicate the index of the sample, and not the parameter value. From this figure, we can see that as the parameter $\sigma$ increases, the 1-cycles from the three circle model disappear and only the primary circle remains for larger values.

\section{Witness Complexes}
\label{witness}

As stated in the introduction, the witness complex construction relies on the selection of a landmark set $L \subset X$. We are interested in determining how the homology of the witness complexes for different landmark selections relate to each other. 

\subsection{Definitions}

In this section, we review some definitions for completeness. A more thorough discussion of witness complexes can be found in \cite{Witness}. 

Let $X$ be a simplicial complex, and denote the $k$-skeleton of $X$ by $X_k$. We let $L$ be a subset of the points of $X$, so $L \subset X_0$, and let $d$ be a metric on the points $X_0$. The set $L$ is referred to as the \emph{landmark set}.

\begin{definition}
Suppose $\sigma$ is a simplex with vertices in $L$. A weak witness for $\sigma$ is a point, $x$, that satisfies
$$d(x, u) \leq d(x, v) \mbox{ for all } u \in \sigma, v \in L \setminus \sigma$$
\end{definition}

In other words, if $\sigma$ is an $n$-simplex, then $x$ is a weak witness for $\sigma$ if the $(n+1)$-nearest neighbors are the vertices of $\sigma$.

\begin{definition}
The weak witness complex $W(X; L)$ consists of simplicies for which there exists a weak witness in $X$, and such that all subsimplices have weak witnesses.
\end{definition}

For example, suppose that in the figure below we have that $L = \{A, B, C, D\}$. Then $E$ is a (weak) witness for the edge $[B, C]$ since the 2 nearest neighbors of $E$ in $L$ are $B$ and $C$. The edge $[B,D]$ is not in the weak witness complex since there is no point outside of $L$ that witnesses $B$ and $D$ (in other words $B$ and $D$ are not the two nearest neighbors of any point outside $L$). 

\vspace{0.5cm}
\hspace{6cm}\includegraphics[height=4cm]{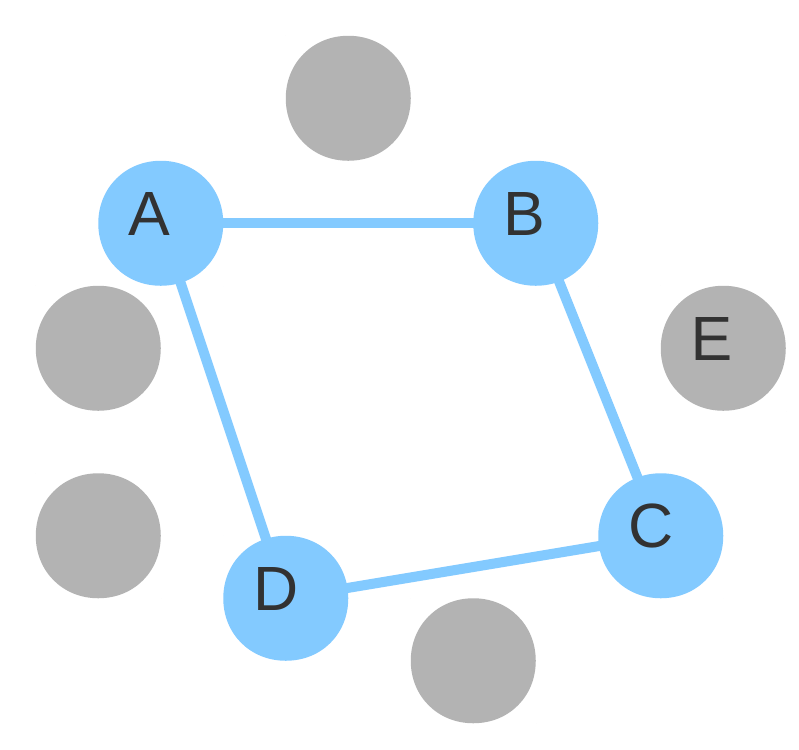}
\vspace{0.5cm}

It is important to note that in general the witness construction is not functorial in that if $L \subset L'$, it is not necessarily true that $W(X; L) \subset W(X; L')$. If this were true, one would be able to compare different landmark selections by forming a zigzag complex of topological spaces with terms $W(X; L) \hookrightarrow W(X; L \cup L') \hookleftarrow W(X; L')$. Due to the failure of this property, we must use an alternate construction known as the witness bicomplex. For the proceeding definitions, suppose we have subsets $L, M \subset X_0$.

\begin{definition}
$x$ in $X$ is a weak biwitness for the bisimplex $(\sigma, \tau)$ if $x$ is a weak witness for $\sigma$ in $L$ and a weak  witness for $\tau$ in $M$.
\end{definition}

\begin{definition}
The pair $(\sigma, \tau)$ is in the weak biwitness complex $W(X; L, M)$ if all subcells of $(\sigma, \tau)$ have a weak biwitness $x \in X$.
\end{definition}
Note that elements of this complex are pairs $(\sigma, \tau)$ with the vertices of $\sigma$ in $L$ and the vertices of $\tau$ in $M$. We can define the obvious projections
$$p_1: W(X; L, M) \rightarrow W(X; L)$$
$$p_2: W(X; L, M) \rightarrow W(X; M)$$
with $p_1: (\sigma, \tau) \mapsto \sigma$ and $p_2: (\sigma, \tau) \mapsto \tau$. These are well defined, since the definitions imply that we must have that
$$W(X; L, M) \subset W(X; L) \times W(X; M)$$

The above definitions allow us to consider the following construction. We piece together all of these maps to form what we define as the \emph{biwitness zigzag complex}:
$$W(X; L_0) \leftarrow W(X; L_0, L_1) \rightarrow  W(X; L_1) \leftarrow W(X; L_1, L_2) \rightarrow \ldots$$



\subsection{A Note on the Selection of the Landmark Points}

In practice, if one is computing persistent homology using a witness complex, the first stage of the process is the selection of the landmark set. While it is possible that a practitioner may have some a priori idea of which points to select, there are two automated methods for generating landmark points. The first method is just to select $|L|$ points uniformly at random. The second is to perform a sequential max-min procedure. To do this, an initial point $L_0 = \{l_0\}$ is selected. Subsequent points are selected by maximizing the minimum distance to the existing landmark set. In other words,
$$L_{n+1} = L_n \cup \left( \arg \max_l \{ \min d(l, l_i) : l_i \in L_n \} \right)$$

In the figure below, a randomized selection is shown on the left, and a max-min selection is shown on the right. Note that the max-min selection produces points that are more evenly spaced. We denote the landmark points in $L$ by black cubes.

\hspace{0cm}\includegraphics[height=7cm]{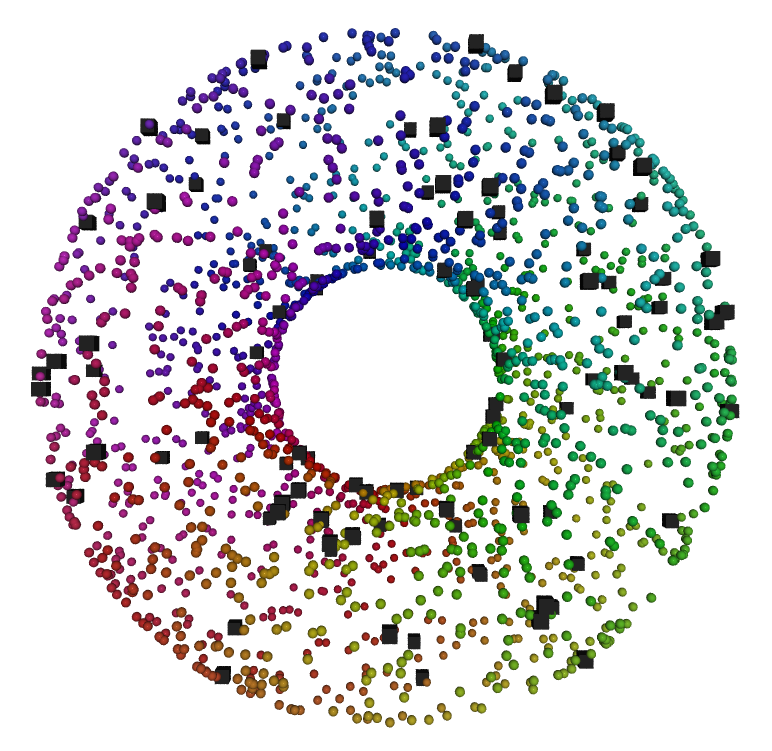}\hspace{1cm}\includegraphics[height=7cm]{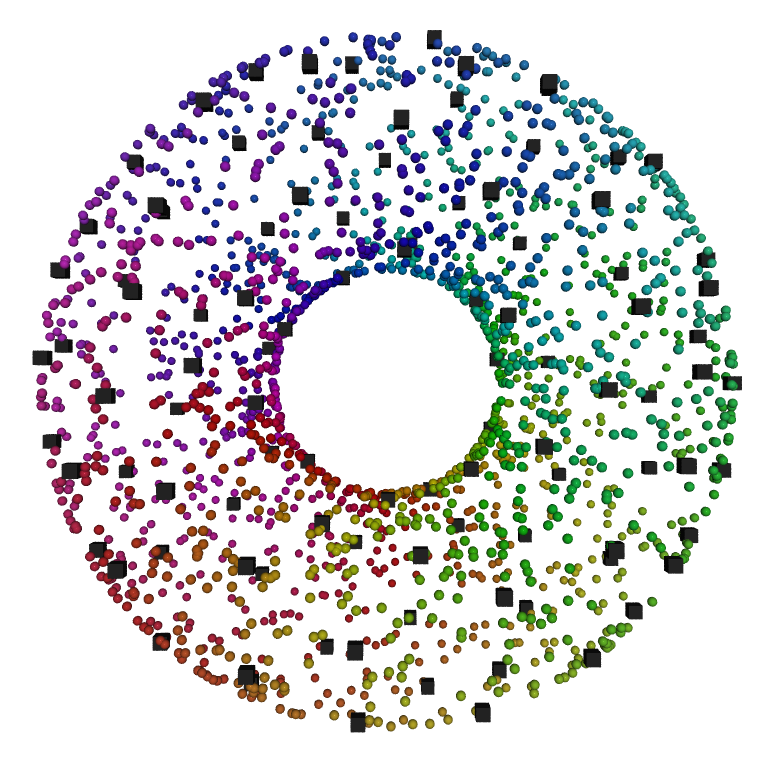}

In this paper we only consider randomized landmark selection. For example, when we show the repeated sampling examples below, all of the selections are randomized. The reason for this is that the only source of randomness in the sequential max-min procedure is in the selction of the first point. When one computes two such landmark sets, very often they share most points with each other. For this reason, the sampling results presented later on would not be very interesting with max-min sampling.

\subsection{Algebraic Formulation}

Let us analyze the algebraic situation of the biwitness zigzag complex. For more background information on related constructions, we refer the reader to \cite{Weibel} or \cite{Lang}. Consider the more general zigzag diagram of differential graded vector spaces (equivalently chain complexes), which we call the \emph{projection diagram}:
$$\cdots \rightarrow A_* \leftarrow C_* \rightarrow B_* \leftarrow \cdots$$
where we have that $C_*$ is a summand of the tensor product $(A \otimes B)_* = \bigoplus_{p+q=*} A_p \otimes B_q$.  We are interested in the action of the projection maps $\pi^1_n$ and $\pi^2_n$ on homology, where we define $\pi^1_n$ and $\pi^2_n$ to be the morphisms:
$$\pi^1_n \left(\sum_j c_j \alpha_j \otimes \beta_j \right) = \sum_j c_j \pi_n(\alpha_j)$$
$$\pi^2_n \left(\sum_j c_j \alpha_j \otimes \beta_j \right) = \sum_j c_j \pi_n(\beta_j)$$
where $\pi_n$ is the (graded) endomorphism which is the identity on grade $n$, and zero elsewhere.

It is straightforward to verify that the morphisms $\pi^i_*$ commute with the boundary operator. To see this, we use the fact that $\pi_n$ trivially commutes with $\partial$ and compute (in the case of $\pi^1_*$ -- the case of $\pi^2_*$ is similar):
\begin{eqnarray*}
\pi^1_{n-1} \partial (\sum_j c_j \alpha_j \otimes \beta_j) & = & \pi^1_{n-1} (\sum_j c_j (\partial \alpha_j \otimes \beta_j + (-1)^{|\alpha_j|} \alpha_j \otimes \partial \beta_j))\\
 & = & \sum_j c_j(\pi_{n-1} (\partial \alpha_j) + (-1)^{|\alpha_j|} \lambda_{\beta_j} \pi_{n-1}(\alpha_j))\\
 & = & \sum_j c_j \pi_{n-1} (\partial \alpha_j)\\
 & = & \sum_j c_j \partial \pi_{n} (\alpha_j)\\
 & = & \partial \pi^1_n(\sum_j c_j \alpha_j \otimes \beta_j)
\end{eqnarray*}
In the above, $\lambda_{\beta}$ denotes the sum of the coefficients in the chain $\partial \beta$. If we were to have that the second term in the second line would be non-zero, then we must have that the degree of $\alpha_j$ must be $n-1$ in order to not get sent to zero by $\pi_{n-1}$. Since $|\alpha_j| + |\beta_j| = n$ we must have that $|\beta_j| = 1$. However, in all cases of interest (for example simplicial or cellular homology) the sum of the coefficients of the boundary of 1-dimensional elements is always zero, so we always have that $\lambda_{\beta_j} \pi_{n-1}(\alpha_j) = 0$. 

Thus, we have that $\pi^i_n$ descends to a map on homology. As in section \ref{resampling} it is possible to compute the interval decomposition of the projection diagram by looking at projections of homology classes. In section \ref{witnessalgsection} we use this observation to build an algorithm analogous to algorithm \ref{unionalg2}.


\subsection{Algorithm}
\label{witnessalgsection}

In order to understand how the homology of different landmark samples relate to each other, we apply the functor $\Ho_p(-)$ to the witness bicomplex diagram, to get a projection diagram:
$$\Ho_p(W(X; L_0)) \leftarrow \Ho_p(W(X; L_0, L_1)) \rightarrow \Ho_p(W(X; L_1)) \leftarrow \Ho_p(W(X; L_1, L_2)) \rightarrow \ldots$$
Let us index the terms in the above diagram as follows:
$$\Ho_p(W(X; L_j)) \leftrightarrow j$$
$$\Ho_p(W(X; L_j, L_{j+1})) \leftrightarrow j + \frac{1}{2}$$
Thus the regular witness complexes are at integral indices, whereas the bicomplexes are at indices of the form $j + \frac{1}{2}$.

In this section we describe the algorithm for computing the interval decomposition of the witness projection diagram. The algorithm incrementally builds up the multi-set of intervals $\mathcal{I}_j$, starting with $\mathcal{I}_0$ containing intervals of the form $[0, \infty]$ for each ``active'' generator of $\Ho_p(W(X; L_0))$. Note that this itself relies on the computation of the (persistent) homology of the individual witness complexes (see \cite{Carlsson_04}). A homology class is said to be active if it corresponds to an infinite interval within the persistent homology $\Ho_p(W(X; L_j))$. We also note that the algorithm described below ignores those intervals that correspond to the biwitness objects within the witness bicomplex diagram. In other words, all of the intervals produces have integral endpoints. In the algorithm, we use the notation $\sim$ to denote the equivalence relation that two cycles differ by a boundary. We are now ready to describe the procedure in algorithm \ref{witnessalg}.

\begin{algorithm}
\caption{Interval decomposition of bicomplex zigzag}
\label{witnessalg}
\begin{algorithmic}

\STATE Initialize $\mathcal{I}_0 \gets \{I_a = [0, \infty] : \mbox{ $a$ is an active homology class in $\Ho_p(W(X; L_0))$} \}$

\FOR{$j = 1, \ldots, n-1$}

\STATE $\mathcal{I}_j \gets \mathcal{I}_{j-1}$

\FOR{$c \in \Ho_p(W(X; L_j, L_{j+1}))$} 

\STATE $c_1 \gets \pi^1_p(c)$\\
\STATE $c_2 \gets \pi^2_p(c)$\\
\IF {$\exists a \in \Ho_p(W(X; L_j)) \mbox{ and } b \in \Ho_p(W(X; L_{j+1})) \mbox{ such that } a \sim c_1 \mbox{ and } b \sim c_2$}
\STATE Maintain the interval $I_a = [s_a, \infty]$ corresponding to the homology class $a$ in $\mathcal{I}_j$
\STATE Mark homology classes $a$ and $b$ as matched
\ENDIF

\ENDFOR

\FOR{$a \in \Ho_p(W(X; L_j)) \mbox{ such that $a$ is unmatched}$}
\STATE End the interval corresponding to the class $a$ by changing $I_a = [s_a, \infty]$ to $I_a \gets [s_a, j]$
\ENDFOR

\FOR{$b \in \Ho_p(W(X; L_{j+1})) \mbox{ such that $b$ is unmatched}$} 
\STATE Start an interval corresponding to $b$ by adding $I_b = [j, \infty]$ to $\mathcal{I}_j$
\ENDFOR

\ENDFOR

\STATE Close off all remaining intervals of the form $I_a = [s_a, \infty]$ by setting them to $I \gets [s_a, n]$

\end{algorithmic}
\end{algorithm}

\subsection{Interpretation}

Suppose we find in the interval decomposition, an interval of the form $[m, m+1]$ where $m$ is an integer. This means that there is a $p$-dimensional homology class that is present in the witness complexes at indices $m$ and $m + 1$, and furthermore there is a homology class in $W(X; L_m, L_{m+1})$ that projects to both of these. Similarly, an interval of the form $[m, n]$ with $m$ and $n$ integers indicates the presence of a homology class that is mapped to by the intermediate bicomplexes for indices $j = m, \ldots, n$. In general, there will be also intervals with half-integer endpoints. These correspond to classes that are born or die at the bicomplexes. Depending on the application in mind, it may be useful to either keep these or discard these. The interval decomposition is a very parsimonious representation of the witness bicomplex zigzag. At any single integer point, the number of ``active'' intervals indicates the rank of the homology group at the given dimension of consideration. Continuity of intervals indicates the preservation of homology classes through a bicomplex.

\subsection{Examples}



\subsubsection{A Synthetic Example}

Let us verify our intuition on a simple example found in \cite{Bicomplex}. Nevertheless, this is an actual example of a witness bicomplex sequence, if we consider the points of $X$ to be the points on the unit circle at angles $\{0, 2 \pi / 3, 4 \pi / 3\}$, and the points of $Y$ to be at angles $\{\pi / 3, \pi, 5 \pi / 3\}$. The following figure shows the bicomplex $Z \subset X \times Y$.

\hspace{1cm}\includegraphics[width=14cm]{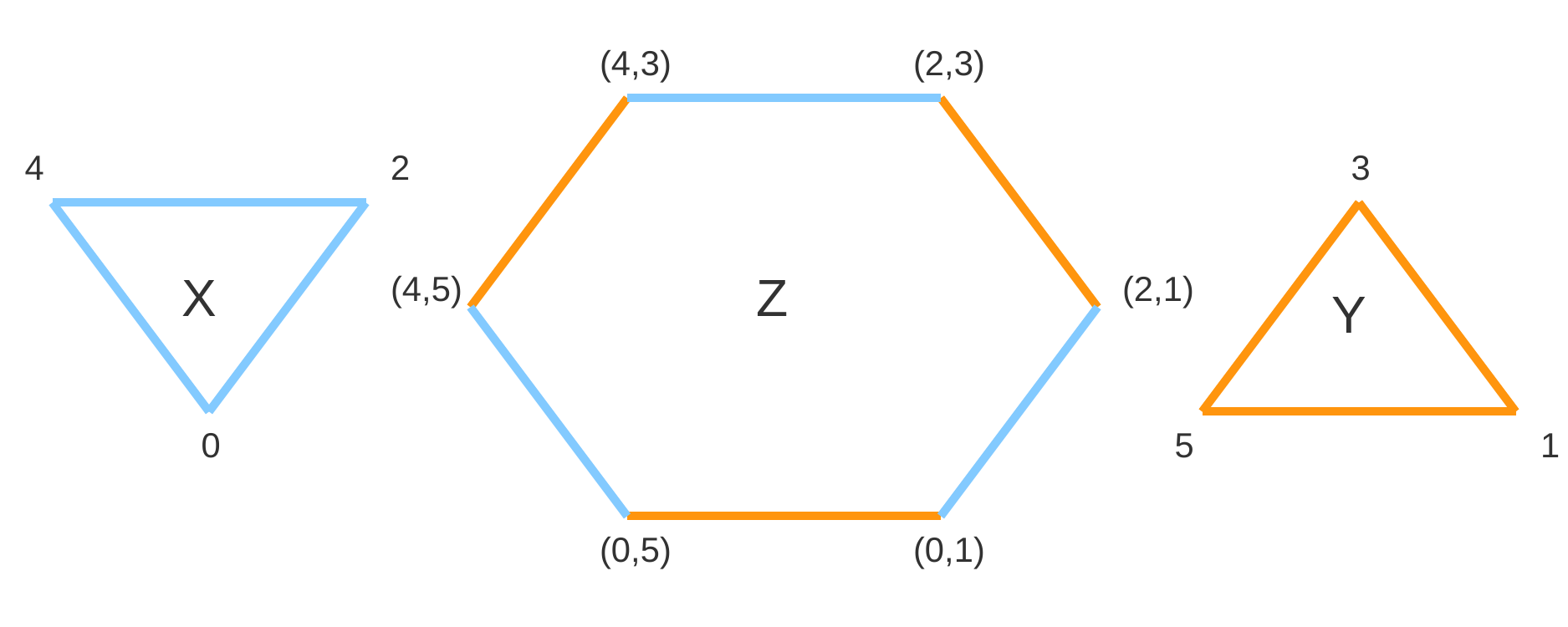}

Another way of thinking about this is $X \times Y$ is a torus cell complex, and $Z$ traces out a loop on its 1-skeleton. Let us consider 1-dimensional homology. The three elements of the sequence each have a single generator, which (using $\mathbb{Z}/2\mathbb{Z}$ coefficients) are $[0,4] + [0,2] + [2,4]$, $([0,4], [5]) + ([4], [3,5]) + ([0,2], [1]) + ([2,4], [3]) + ([0], [1,5]) + ([2], [1,3])$, and $[1,3] + [1,5] + [3,5]$. It is easy to see that the generator for the bicomplex projects to both homology classes at the endpoints. Thus in dimension 1, we have an interval $[0, 1]$. Similarly, in dimension 0 we have generators $[0]$, $([0], [1])$, and $[1]$, thus we also have an interval $[0, 1]$ in dimension 0.

\subsubsection{Verification of Specifically Chosen Landmark Points}

In the figure below, we show three landmark selections on a figure-8. Note that they were constructed so that $C = A \coprod B$. Computing the interval decomposition for the homology of the sequence
$$W(X; A) \leftarrow W(X; A, B) \rightarrow W(X; B)$$
we get the intervals $\{[0, 1]\}$ in dimension 0, and $\{[0, 0], [1, 1]\}$ in dimension 1. In other words, the interval decomposition shows us that the homology classes in $A$ and $B$ in dimension 1 are not compatible. If we do the same thing for the pair $A, C$ we get the intervals $\{[0, 1]\}$ in dimension 0, and $\{[0, 1], [1, 1]\}$ in dimension 1, which shows us the compatibility of a 1-dimensional homology class between $A$ and $C$ and the presence of an isolated one in $C$. Similarly, we get the same result for the pair $B, C$. 

It is important to note that in these examples, we have not included those intervals that start or end at half-integer indices. The reason for this is that the witness bicomplexes can be quite complicated in general, and including such intervals would obscure the intervals of interest -- that is the ones that indicate the mutual projection to homology classes in different integer indices.

\vspace{0.5cm}
\hspace{0.2cm}\includegraphics[width=5cm]{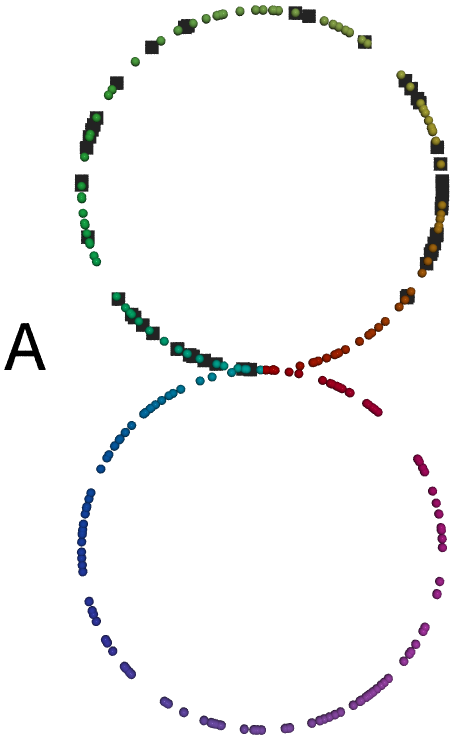}\includegraphics[width=5cm]{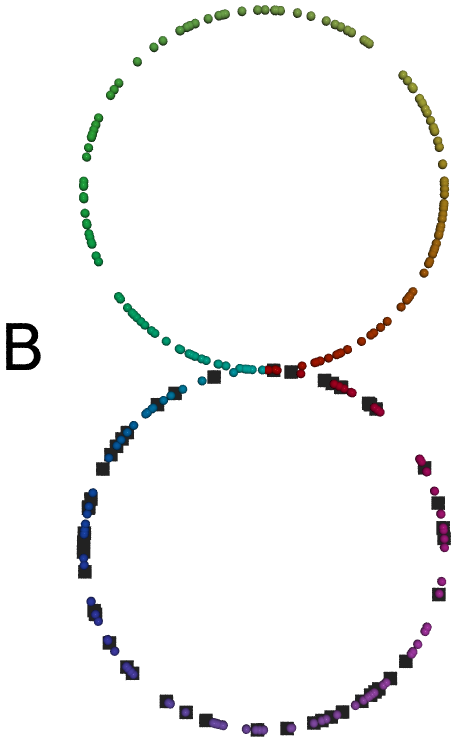}\includegraphics[width=5cm]{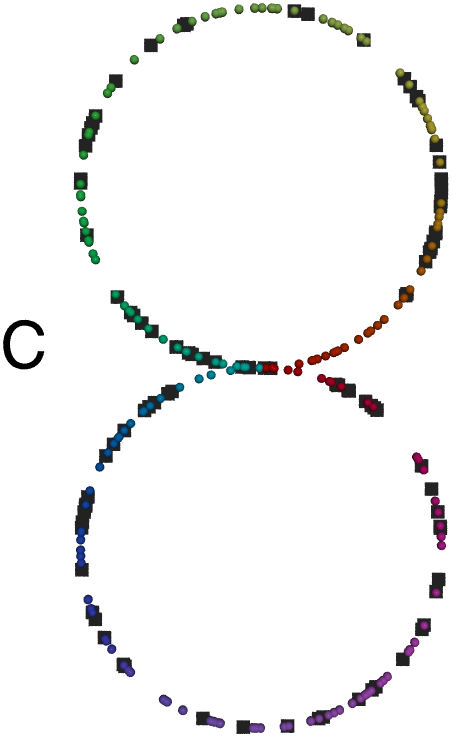}
\vspace{0.5cm}

\subsubsection{Long Term Behavior}
\label{longterm}

In the previous two subsections, we showed examples with only two terms. The next example shows the intervals for 1000 random landmark samples each of size 20, drawn from a point cloud which consists of 1000 randomly selected points on the unit circle. From the figures below, we can see that there are long intervals in dimension 1, yet there is not one continuous interval. This is due to the fact that it is possible for two different landmark selections to produce 1-cycles that are not homologous within the bicomplex. This is expected since the sampling of both the points on the circle as well as the landmark points are done randomly. 

\vspace{0.5cm}
\hspace{3cm}\includegraphics[width=10cm]{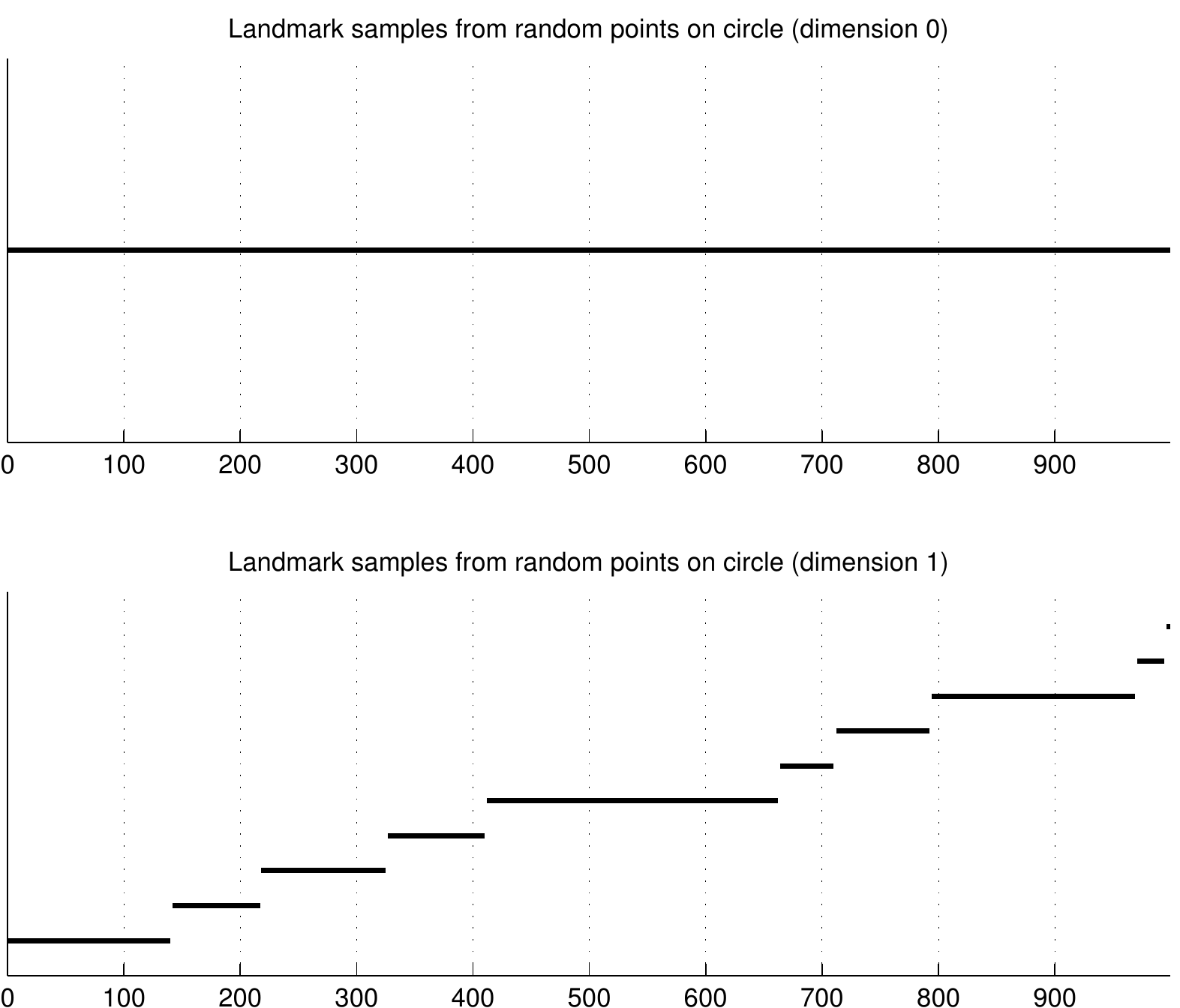}
\vspace{0.5cm}

We can repeat the above using randomly sampled points on a wedge sum of two circles. For the figure below, 41 samples of 40 points were selected from 1000 random points on a figure-8. We only show the intervals in dimension 1, since in dimension 0 there is just 1 long interval as in the circle example. It is interesting to see that although at any point there are two active intervals (corresponding to the two 1-cycles), the samples do not always match up.

\vspace{0.5cm}
\hspace{3cm}\includegraphics[width=10cm]{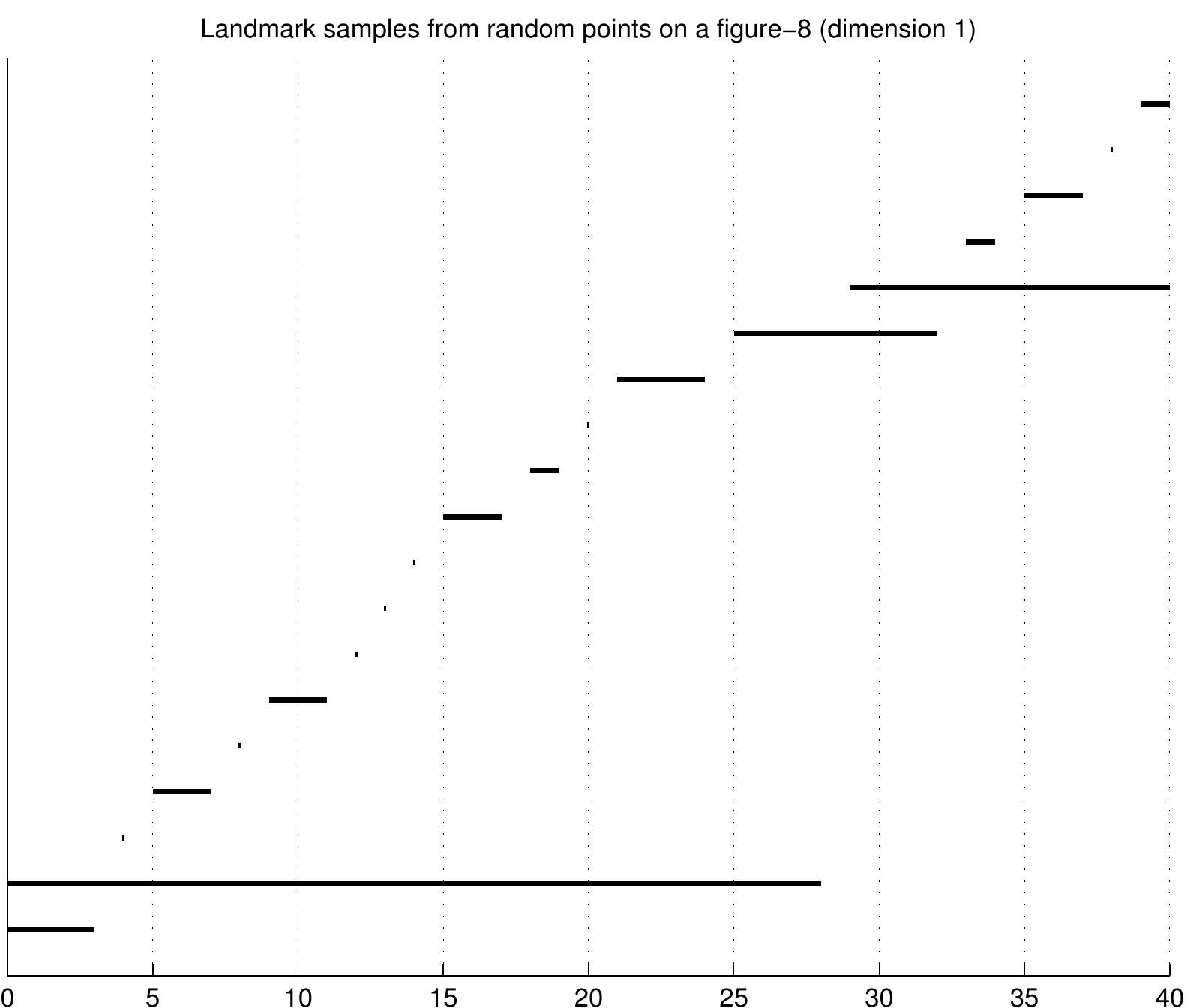}
\vspace{0.5cm}

Additionally, if one is interested in doing a more detailed analysis of the compatibility between the two, it is possible to select a subset of the original set of samples which were included in ``long'' intervals. Through this process, a practitioner may amplify the homological signal obtained in certain samples. In the figure below, this was performed with the samples at indices 6, 7, 10, 15, 16, 17, 21, 22, 23, 24, 25, 26 and 27.

\vspace{0.5cm}
\hspace{3cm}\includegraphics[width=10cm]{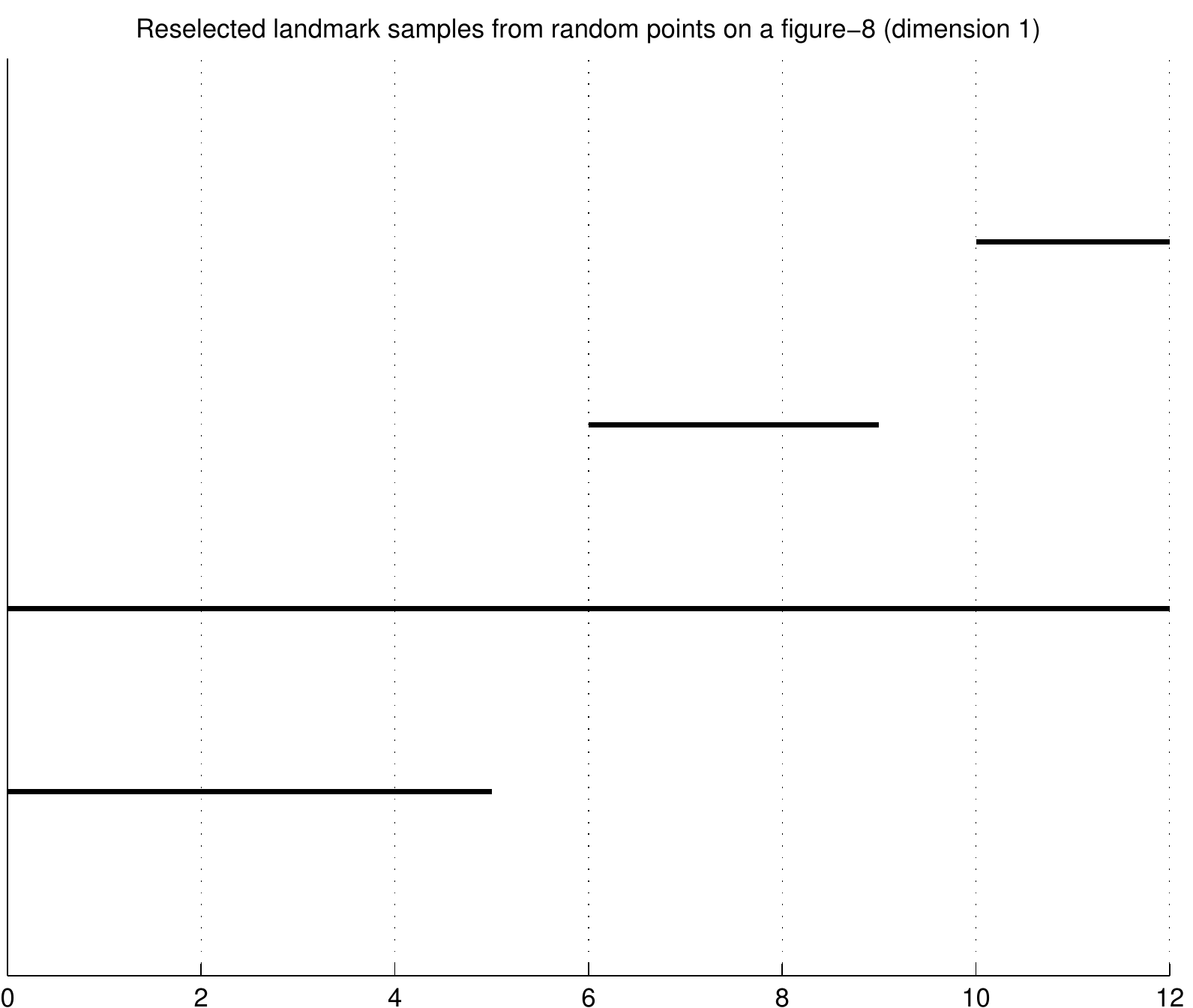}
\vspace{0.5cm}

Another example is the 2-torus constructed by taking $S^1 \times S^1 \subset \mathbb{R}^4$. We take 10,000 points uniformly at random, and construct 40 samples of 50 points. Furthermore, we only select those landmark choices for which the Betti numbers of the witness complexes are $\{1, 2, 1\}$. The interval plots for the zigzag homology of the bicomplex diagram are:

\vspace{0.5cm}
\hspace{3cm}\includegraphics[width=10cm]{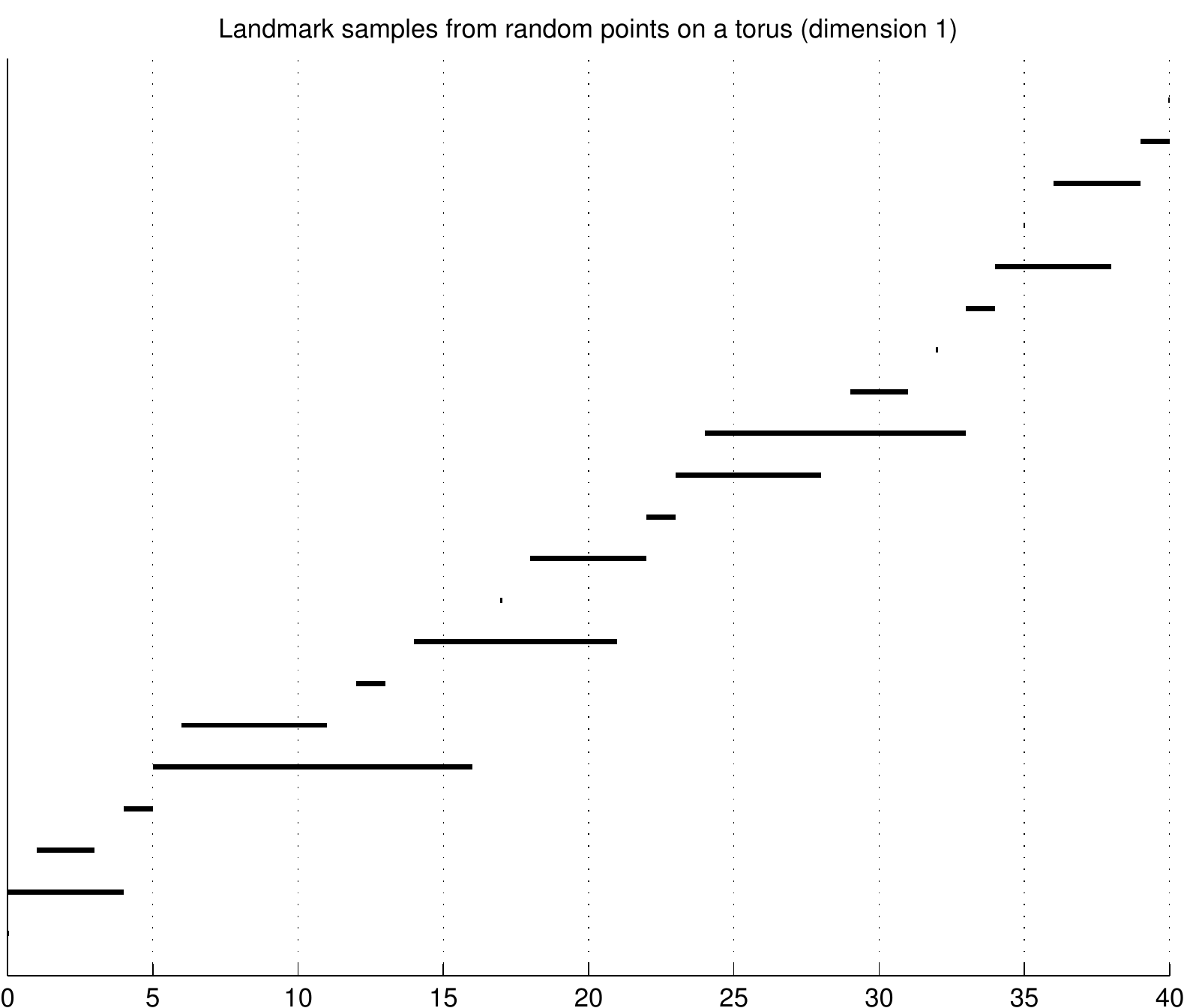}
\vspace{0.5cm}

\hspace{3cm}\includegraphics[width=10cm]{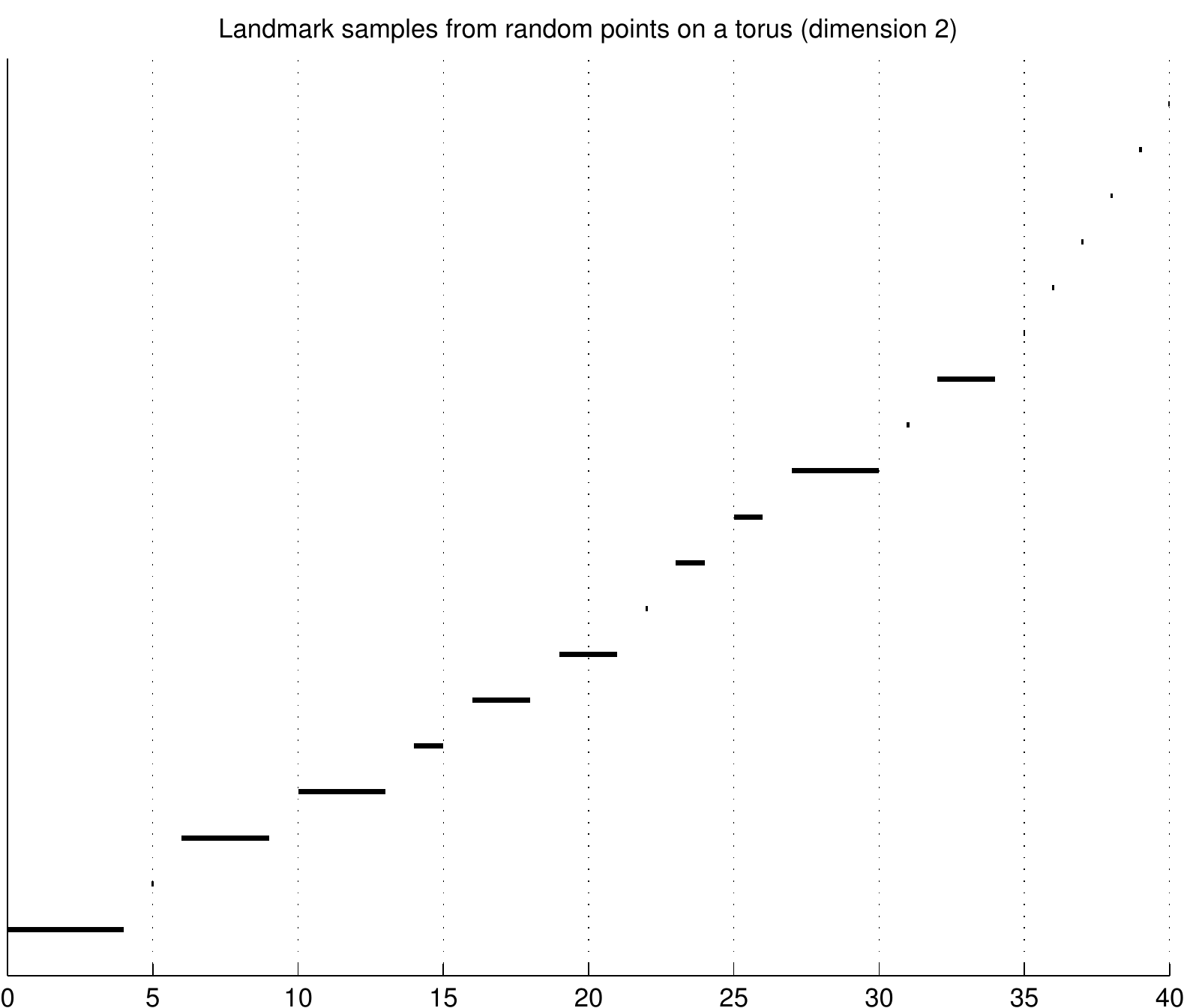}
\vspace{0.5cm}

\subsubsection{Incremental Behavior}

One interesting question to ask is how different size landmark selections relate to each other. Using the witness bicomplex framework, this is easy to investigate. We select a list of sizes $\{n_0, \ldots, n_k\}$ and generate landmark selections $\{L_i\}$ each with $n_i$ points. The next example shows this where the underlying dataset is a random sample of 200 points on the unit circle, and the sizes are $\{2, 3, 4, \ldots, 80\}$

\vspace{0.5cm}
\hspace{3cm}\includegraphics[width=10cm]{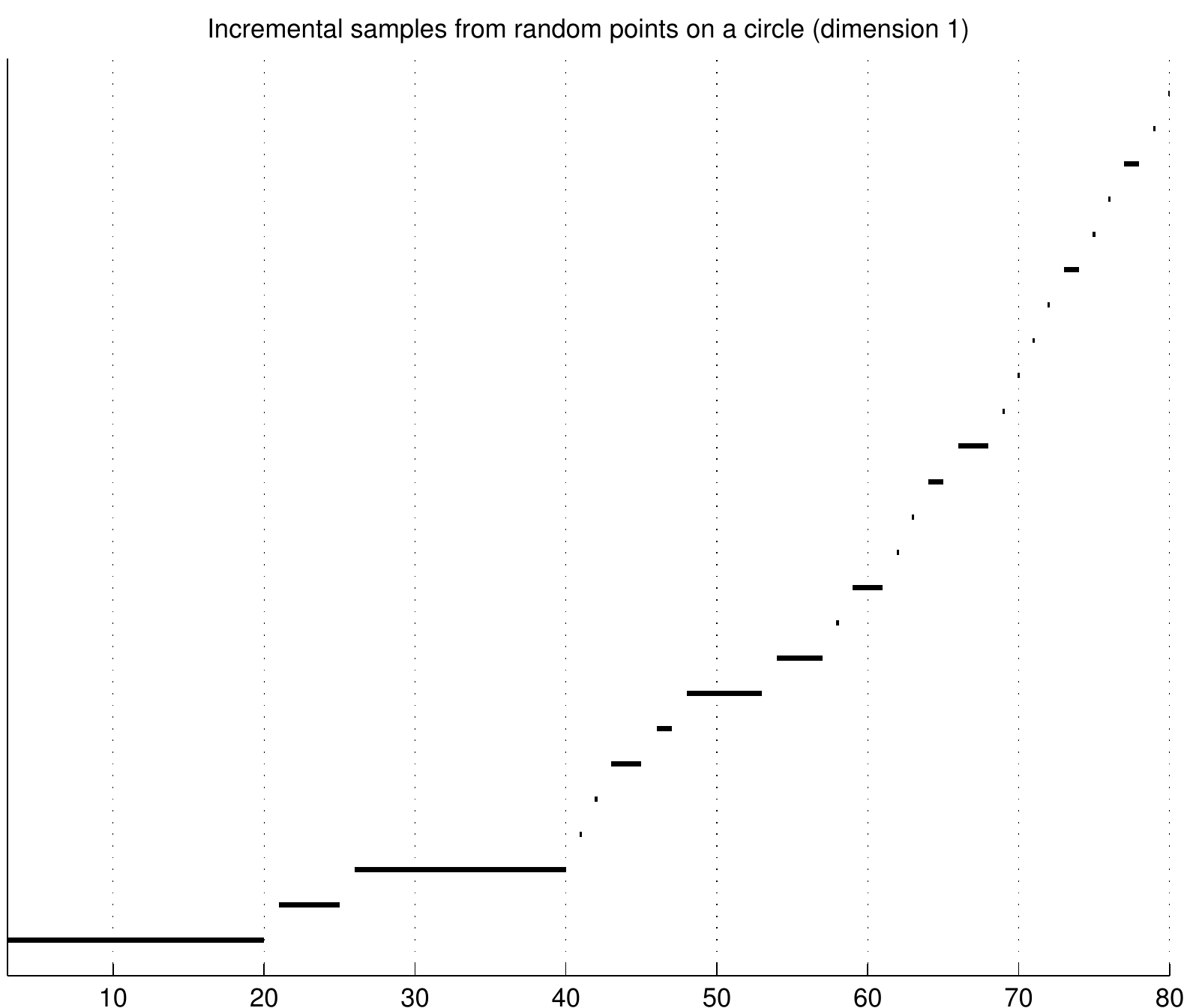}
\vspace{0.5cm}

The reason for the long intervals among small landmark set sizes is that in order for an interval to continue between two integer indices (two witness complexes), we need there to be sufficiently many witness in the complement of the landmark selection that ``connect'' the two homology classes in the different selections. As the size of the landmark sets increase, this criteria (the existence of the appropriate witnesses) fails to be satisfied. Similarly, for the figure-8, using the same parameters we get the following:

\vspace{0.5cm}
\hspace{3cm}\includegraphics[width=10cm]{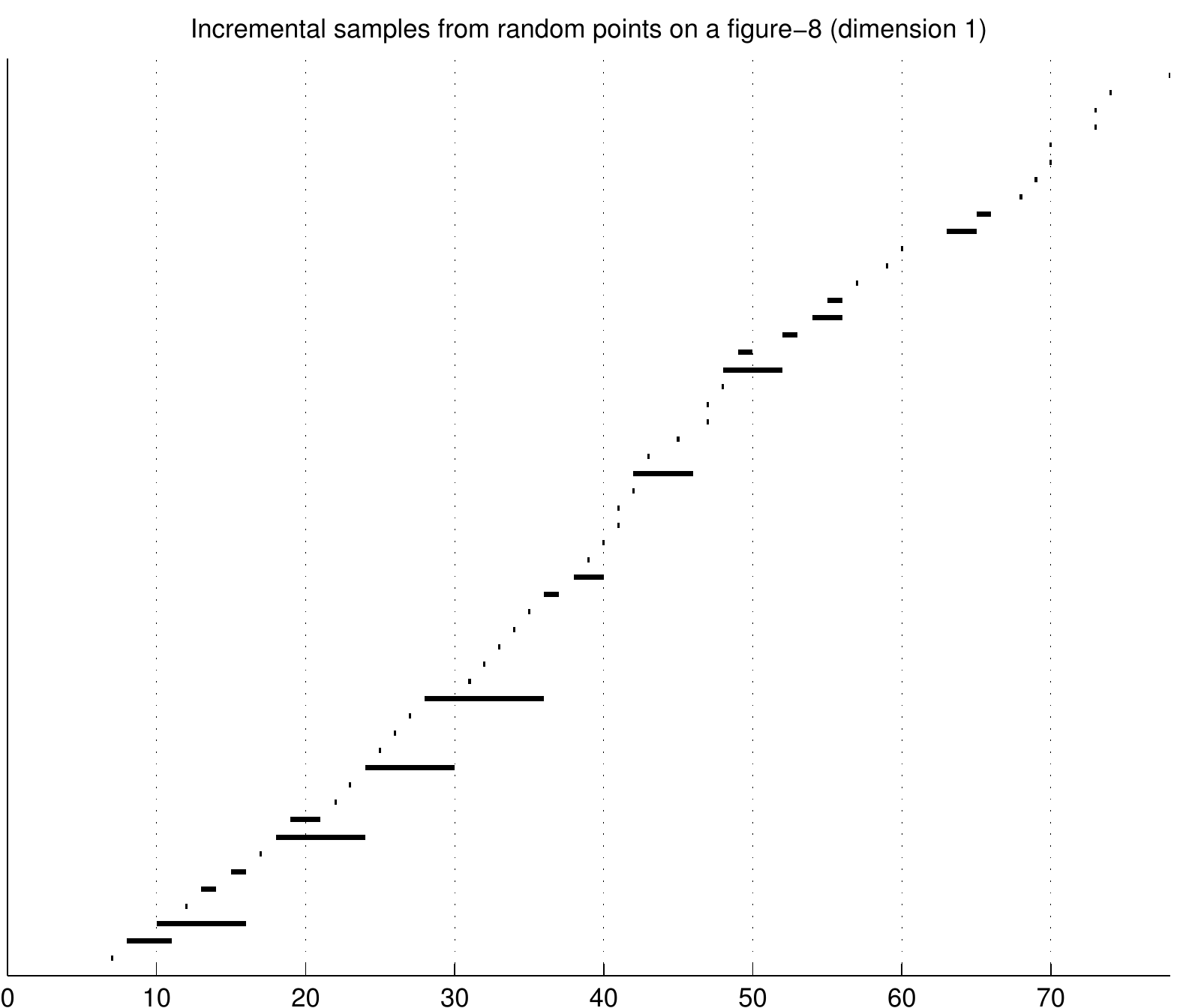}
\vspace{0.5cm}

\subsubsection{Pairwise Comparison}

Up until this point, we have avoided talking about the ordering of the terms within the zigzag diagram. It is obvious that one may get different interval decompositions using a different rearrangement of the witness complexes. Instead, one may wonder whether it is possible to do a mutual comparison. To do this one would form a complete graph $K_n$ (if we suppose that we have $n$ landmark selections $\{L_i\}$). Vertex $i$ would correspond to the witness complex $W(X; L_i)$, and the edge $(i, j)$ would correspond to $W(X; L_i) \leftarrow W(X; L_i, L_j) \rightarrow W(X; L_j)$. However, for important reasons the homology of this ``complete'' diagram is not algebraically tractable as in the linear case. The reasons for this are beyond the scope of this paper, but an interested reader may consult \cite{ZigZag1}.

One can approach this instead by performing pairwise comparisons, rather than a mutual comparison. That is, suppose we have $n$ landmark selections, we form all unordered pairs $(i, j) \subset \{1, \ldots, n\}$ and compute the interval decomposition of the zigzag diagram involving only the selections $L_i$ and $L_j$ at one time. 

For the example below, we perform this procedure as follows. For each distinct pair $(i, j)$, we compute the homology of the diagram $W(X; L_i) \leftarrow W(X; L_i, L_j) \rightarrow W(X; L_j)$. Our underlying set is a 1000 point sample from a figure-8. For visual purposes, we construct a graph on $n$ vertices where we connect the edge $(i, j)$ if the interval decomposition is $\{[0, 1]\}$ in dimension 1 and $\{[0, 1], [0, 1]\}$ in dimension 1. The vertices correspond to different random landmark selections. For the figure below, we used 41 samples each of size 40. 

\vspace{0.5cm}
\hspace{3cm}\includegraphics[width=10cm]{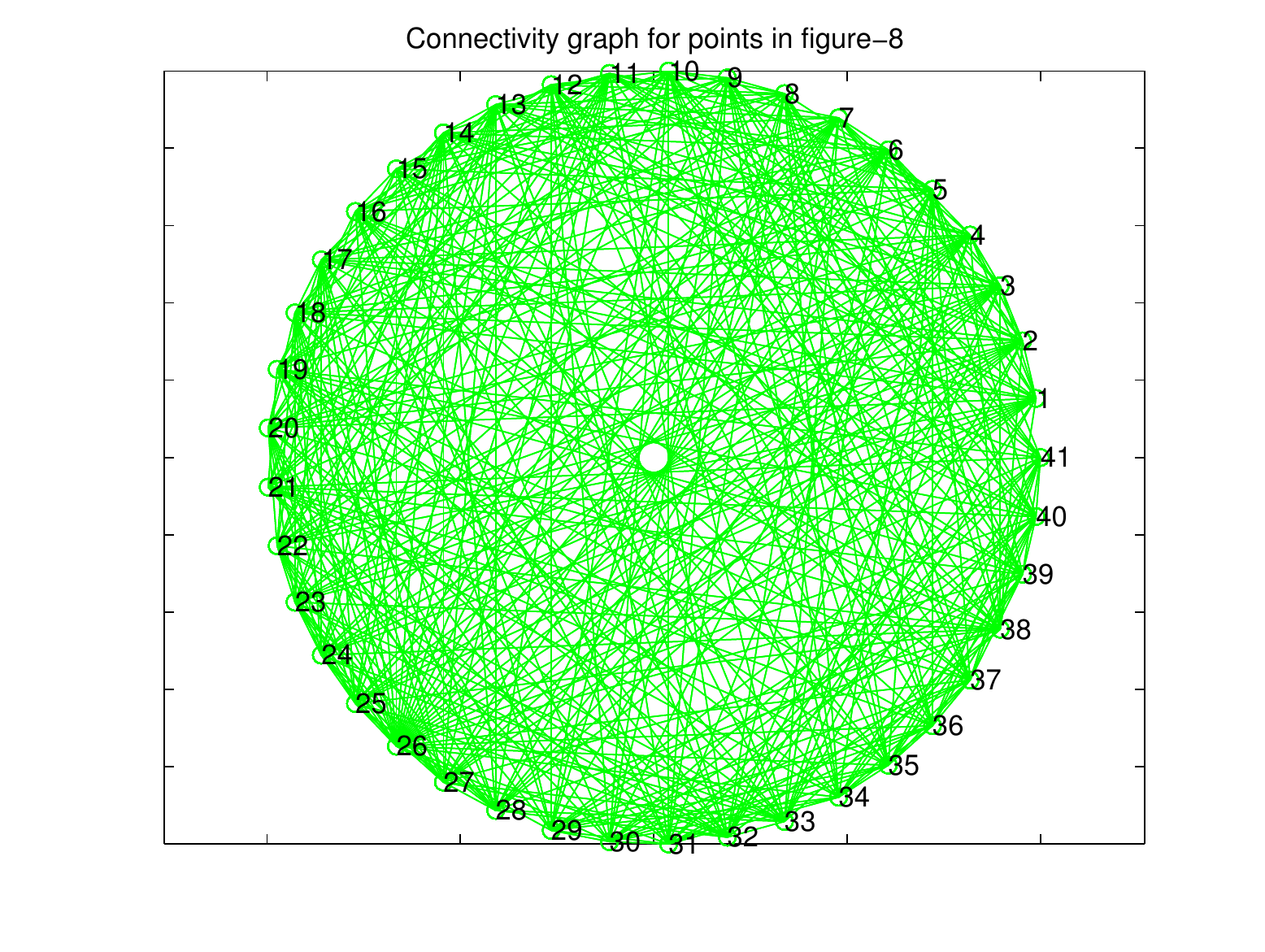}
\vspace{0.5cm}

Although the above graph is too dense to have much visual meaning, it contains all of the possible compatibility information we wish to know between each sample. 



\subsubsection{Image Patch Data}

In this section we consider a much more realistic example, namely the image patch dataset considered in \cite{Images} and \cite{Lee}. As discussed in section \ref{impatch1}, the set $\mathcal{M}_0$ consists of $5 \times 10^4$ high-contrast patches taken from a database of natural images. In \cite{Images} and \cite{Carlsson_09}, using witness complexes the authors extract both the primary circle and secondary circle models. Using the above witness complex comparison framework, it is possible to investigate the role of the landmark sets.



We before, we use the $k$-codensity function:
$$\delta_k(x) = d(x, \nu_k(x))$$
where $\nu_k(x)$ is the $k$-th nearest neighbor of the point $x$. The interpretation is that $\delta_k$ is inversely proportional to the density of the dataset. We may perform the witness complex sampling procedure to verify that there is indeed one primary circle in $\mathcal{M}_{0,\delta}[300, 30\%]$. We take 50 samples each of size 30, and get the following intervals:

\vspace{0.5cm}
\hspace{3cm}\includegraphics[width=10cm]{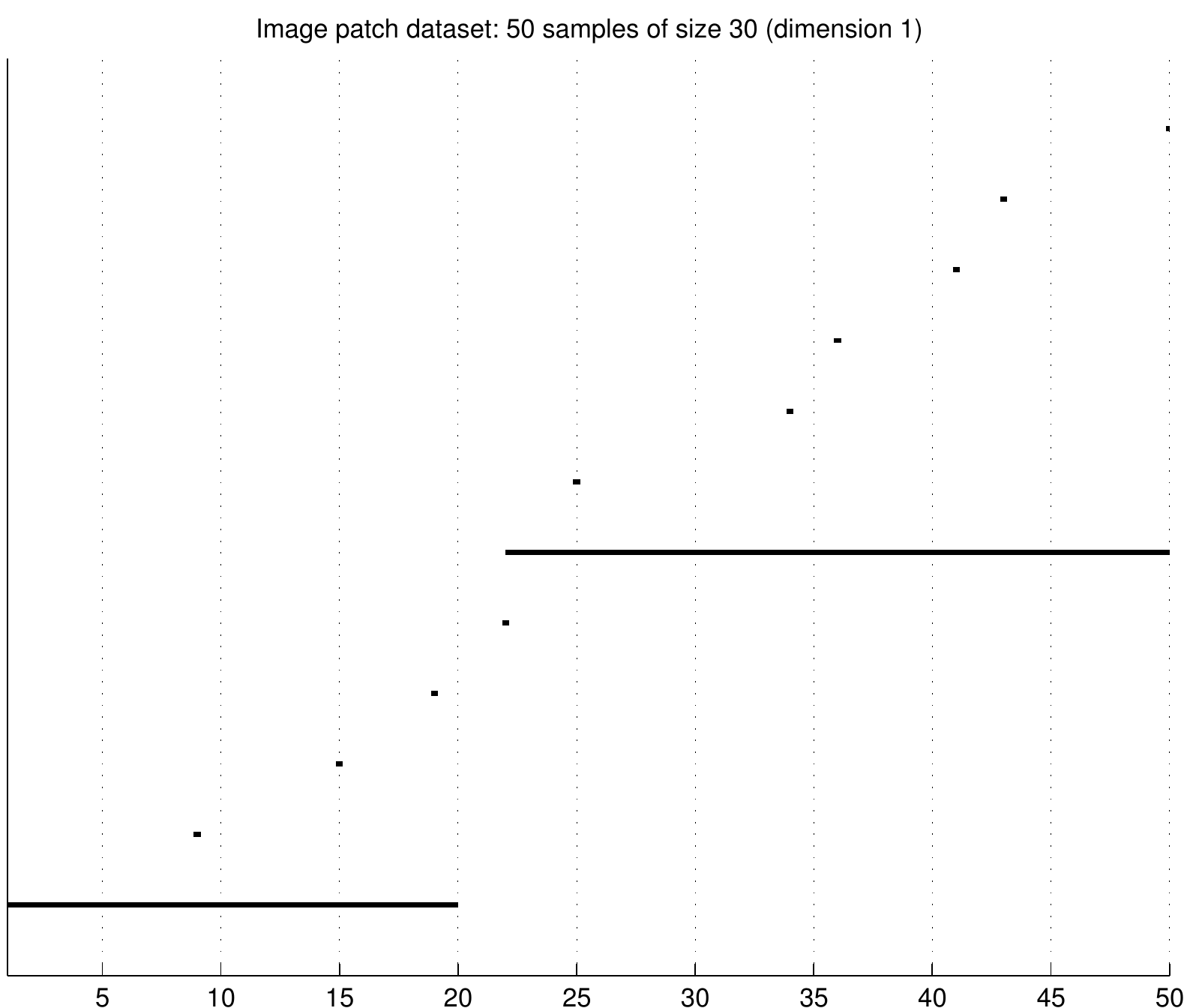}
\vspace{0.5cm}

Note that there are certain samples in which there is no 1-dimensional homology class and there are samples in which there are more than one. However, if we force our samples to produce a 1-cycle (by rejecting those samples that do not have exactly one), we get the following set of intervals:

\vspace{0.5cm}
\hspace{3cm}\includegraphics[width=10cm]{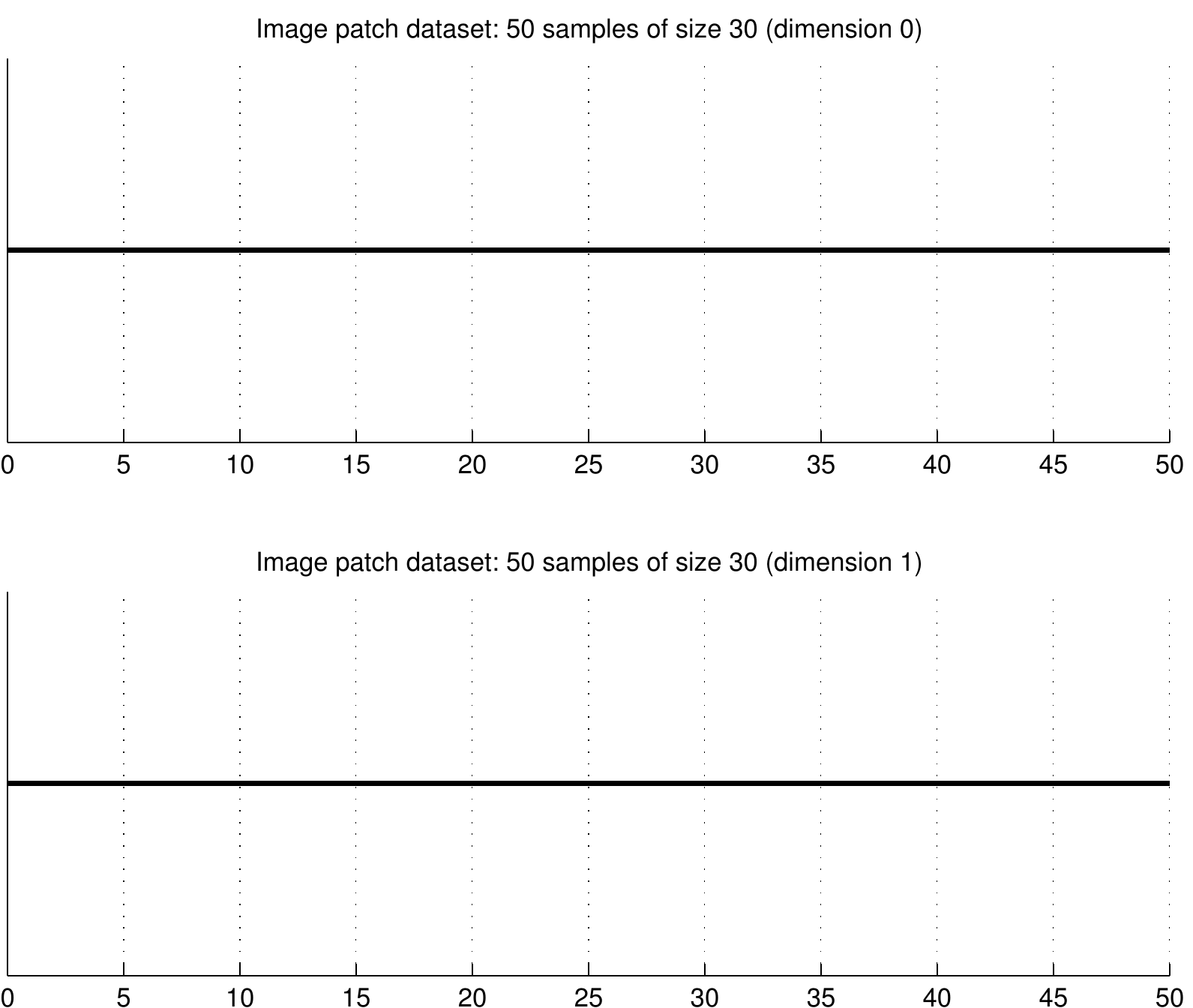}
\vspace{0.5cm}

The above figure tells us that all of the samples that measure a 1-dimensional cycle are compatible, supporting the conclusion in \cite{Adams} that there is one primary circle at this density sampling. At this point the reader may wonder why the above result is so clean, yet in previous synthetic examples (for example the first one in section \ref{longterm}) there are some points of discontinuity. There are two reasons for this. First, the above set of intervals was constructed by selecting only those samples in which a 1-cycle appeared. Second, if we consider the first example of section \ref{longterm} then we can see that out of 1000 samples there were only 9 points of discontinuity, meaning that they are relatively rare.

\section{Concluding Remarks}

In this paper we have demonstrated the use of zigzag persistent homology in three applications: topological bootstrapping, the comparison of thresholding parameters, and the comparison of landmark selections for witness complexes. Zigzag persistence allows one to obtain a multi-set of intervals which indicate the preservation of homological features across different samples, levelsets or landmark selections. Long intervals indicate the stability of such features across many terms in the relevant sequence. Additionally, we have shown the use of these techniques in several computational examples and have demonstrated that these methodologies reveal important information about nonlinear datasets.

\section{Acknowledgements}

The authors would like to thank Henry Adams for sharing his expertise and providing assistance with the image patch dataset. 

\bibliographystyle{amsalpha}
\bibliography{biblio}

\end{document}